\documentclass[11pt]{article}
\usepackage[margin=0.9in]{geometry}

\usepackage{rotating}
\usepackage[position=top]{subfig}
\usepackage{pdfpages}
\usepackage{algorithm2e}
\usepackage{amsmath,amssymb,amsthm,amsfonts,algorithmic,enumerate,appendix, color}
\usepackage{booktabs}
\usepackage{tabularx}
\usepackage{multirow}
\usepackage{numprint}
\usepackage{color}
\npdecimalsign{.}
\nprounddigits{2}
\newcommand\measureISpecification{4ex}
\newcommand{\ctab}[1]{\raisebox{\dimexpr \measureISpecification/2 -.748ex}{#1}}
\usepackage[colorlinks=true,linkcolor=blue,CJKbookmarks=true]{hyperref}
\usepackage{graphicx}
\usepackage{apalike}
\usepackage{hyperref, url}
\newcommand{\x}{\textbf{\textit{x}}}
\renewcommand{\a}{\textbf{\textit{a}}}
\renewcommand{\c}{\textbf{\textit{c}}}
\renewcommand{\b}{\textbf{\textit{b}}}

\SetKwProg{Fn}{Function}{}{}
\SetKwProg{Sr}{Subroutine}{}{}

\makeatletter
\let\@fnsymbol\@arabic
\makeatother
\def \VideoThreePURL{https://www.dropbox.com/s/63j6mayz9wzix8c/3P_PMD.mp4?dl=0}
\def \VideoDendriticURL{https://www.dropbox.com/s/m7r1t9fu62qogef/Dendritic_PMD.mp4?dl=0}
\def \VideoWidefieldURL{https://www.dropbox.com/s/5cdfyxfh315vjz4/Widefield_PMD.mp4?dl=0}
\def \VideoVoltageURL{https://www.dropbox.com/s/xf1a7f1xcckarwb/QState_PMD.mp4?dl=0}
\def \VideoEndoscopeURL{https://www.dropbox.com/s/vfdhoakds9hgzxh/Endoscope_PMD.mp4?dl=0}

\def \VideoEndoscopeBkgURL{https://www.dropbox.com/s/ean0aceif0snhmr/Endoscope_background_PMD.mp4?dl=0}
\def \VideoSuperpixelizationURL{https://drive.google.com/open?id=1VPnDlG2Z0XRrSFl9HEUROl1rsKD15Du5}
\def \VideoDemixVoltageURL{https://drive.google.com/open?id=1kJjZ60QfDwkEKLKwgDhGXgUAgUKclXdO}
\def \VideoDemixDendriticURL{https://drive.google.com/open?id=1BLx2aEBkKCJDO7x-mB6iIY5kLcvrhvH8}
\def \VideoSimulateDendriticURL{https://drive.google.com/open?id=1dPrbxEJftKQ2FuYJ3GHviFZDohoUcKXY}
\title{Penalized matrix decomposition for denoising, compression, and improved demixing of functional imaging data}
\author{E.\ Kelly Buchanan\thanks{Equal contribution, arranged alphabetically; ekb2154, iak2119, dz2336@columbia.edu},$^,$\footnotemark[2] ~ Ian Kinsella\footnotemark[1],$^,$\footnotemark[2] ~ Ding Zhou\footnotemark[1],$^,$\footnotemark[2] ~ Rong Zhu\thanks{Departments of Statistics and Neuroscience, Grossman Center for the Statistics of Mind, Center for Theoretical Neuroscience, and Zuckerman Mind Brain Behavior Institute, Columbia University}, ~ Pengcheng Zhou\footnotemark[2], \\ Felipe Gerhard\thanks{Q-State Biosciences, Inc., Cambridge, MA}, John Ferrante\footnotemark[3],  \\
Ying Ma\thanks{Department of Biomedical Engineering and Zuckerman Mind Brain Behavior Institute, Columbia University}, Sharon H. Kim\footnotemark[4], Mohammed A Shaik\footnotemark[4], \\
Yajie Liang\thanks{Departments of Physics and Molecular and Cell Biology, UC Berkeley}, Rongwen Lu\footnotemark[5], \\ 
Jacob Reimer\thanks{Department of Neuroscience and Center for Neuroscience and Artificial Intelligence, Baylor College of Medicine}, Paul G Fahey\footnotemark[6], Taliah N Muhammad\footnotemark[6], \\
Graham Dempsey\footnotemark[3], Elizabeth Hillman\footnotemark[4], Na Ji\footnotemark[5], Andreas S Tolias\footnotemark[6],  Liam Paninski\footnotemark[2]}
\date{\today}
\begin{document}
\maketitle

\begin{abstract}
Calcium imaging has revolutionized systems neuroscience, providing the ability to image large neural populations with single-cell resolution.  The resulting datasets are quite large (with scales of TB/hour in some cases), which has presented a barrier to routine open sharing of this data, slowing progress in reproducible research.  State of the art methods for analyzing this data are based on non-negative matrix factorization (NMF); these approaches solve a non-convex optimization problem, and are highly effective when good initializations are available, but can break down e.g.~in low-SNR settings where common initialization approaches fail.

Here we introduce an improved approach to compressing and denoising functional imaging data.  The method is based on a spatially-localized penalized matrix decomposition (PMD) of the data to separate (low-dimensional) signal from (temporally-uncorrelated) noise.  This approach can be applied in parallel on local spatial patches and is therefore highly scalable, does not impose non-negativity constraints or require stringent identifiability assumptions (leading to significantly more robust results compared to NMF), and estimates all parameters directly from the data, so no hand-tuning is required.  We have applied the method to a wide range of functional imaging data (including one-photon, two-photon, three-photon, widefield, somatic, axonal, dendritic, calcium, and voltage imaging datasets): in all cases, we observe $\sim$2-4x increases in SNR and compression rates of 20-300x with minimal visible loss of signal, with no adjustment of hyperparameters; this in turn facilitates the process of demixing the observed activity into contributions from individual neurons.  We focus on two challenging applications: dendritic calcium imaging data and voltage imaging data in the context of optogenetic stimulation.  In both cases, we show that our new approach leads to faster and much more robust extraction of activity from the video data.

\end{abstract}
\section*{Introduction}

Functional imaging is a critical tool in neuroscience.  For example, calcium imaging methods are used routinely in hundreds of labs, generating large-scale video datasets whose characteristics (cell shapes, signal-to-noise levels, background activity, signal timescales, etc.) can vary widely depending on the imaging modality and the details of the brain region and cell types being imaged.  To handle this data, scientists must solve two basic tasks: we need to extract signals from the raw video data with minimal noise, and we need to store (and share) the data.  A number of papers have focused on the first task \cite{mukamel2009automated,maruyama2014detecting,pnevmatikakis2016simultaneous,pachitariu2016suite2p,friedrich2017multi,inan2017robust,Reynolds2017,petersen2017scalpel,zhou2018efficient,Mishne2018}; however, somewhat surprisingly, very little work has focused on the second task.

For both of these tasks, it is critical to denoise and compress the data as much as possible.  Boosting the signal-to-noise ratio (SNR) is obviously important for detecting weak signals, performing single-trial analyses (where noise cannot be averaged over multiple trials), and for real-time experiments (where we may need to make decisions based on limited data - i.e., averaging over time is not an option).  The benefits of compression are perhaps less obvious but are just as numerous: compression would facilitate much more widespread, routine open data sharing, enhancing reproducible neuroscience research.  Compression will also be critical for in vivo imaging experiments in untethered animals, where data needs to be transmitted wirelessly, making data bandwidth a critical constraint.  Finally, many signal extraction methods based on matrix factorization can be sped up significantly if run on suitably compressed data.

Previous methods for denoising and compressing functional data have several drawbacks.  Generic video compression approaches do not take advantage of the special structure of functional imaging data and produce visible artifacts at high compression rates; more importantly, these approaches do not denoise the data, since they focus on compressing the full data, including noise, whereas our goal here is to discard the noise.  Conversely, generic image denoising approaches do not offer any compression (and also fail to take advantage of strong structured correlations in the video data). Constrained nonnegative matrix factorization (CNMF) \cite{pnevmatikakis2016simultaneous} approaches provide state of the art denoising and demixing of calcium imaging data, but these methods can leave significant visible signal behind in the residual (discarding potentially valuable signal) and are highly dependent on the initialization of the matrix factorization; thus it would be dangerous to keep only the matrix factorization output and discard the raw data.  Principal components analysis (PCA) is often employed as a compression and denoising method \cite{mukamel2009automated,pachitariu2016suite2p}, but PCA is based on a rather unstructured signal model and therefore provides a suboptimal encoder of functional data (we will discuss this point in further depth below).  In addition, the computation time of PCA scales quadratically with the number of pixels (assuming a long video dataset) and therefore naive applications of PCA are rather slow \cite{friedrich2017multi}.  Finally, importantly, it is difficult to automatically choose the number of principal components that should be retained in a given video (and the “correct” number of components can vary widely across different datasets). 

Here we introduce a new simple approach to denoising and compressing functional video data.  We apply a variant of penalized matrix decomposition \cite{Witten2009pmd} that operates locally in space, and encourages smoothness in both the spatial and temporal dimensions.  This method offers multiple advantages over previous approaches.  It is based on a signal model that is well-matched to the structure of the data: cells are local in space, there aren't too many of them compared to the number of pixels (leading to a low-rank signal model), and cellular activity is smoother than the dominant noise sources, which are spatially and temporally uncorrelated.  The approach is scalable (scaling linearly in the number of frames and pixels), and has modest memory requirements (because all processing is only performed in local spatial patches).  All parameters (including the local matrix rank and the degree of smoothness of the output) are chosen automatically.  Empirically we find that the method is highly effective, leaving behind minimal visible structure in the residual, while achieving 20-300x compression rates and 2-4x improvements in SNR.  We demonstrate the method's effectiveness on a wide variety of functional imaging datasets (both calcium and voltage imaging; one-, two- and three-photon imaging; and data including somas and dendrites) and show that the method is also effective on wide-field imaging data, where single-cell resolution is not available.  Finally, we develop a new constrained NMF approach based on the denoised and compressed representation of the data, and apply this new demixing method to two challenging applications: dendritic calcium imaging data and voltage imaging data in the context of optogenetic stimulation.  In both cases, we show that our new approach leads to faster and much more robust extraction of activity from the video data.

\section*{Methods}

We begin by defining notation. Our starting point is an imaging dataset that has been motion-corrected (i.e., we assume that there is no motion of visible cellular components from frame to frame of the movie) and then ``unfolded" into a $d \times T$ matrix $\mathbf{Y}$, where $T$ is the number of frames in the movie and $d$ is the number of pixels per frame (or voxels per frame if we are performing imaging in three dimensions).  Now the typical approach is to model the data $\mathbf{Y}$ as  $\mathbf{Y} = \mathbf{AC} + \mathbf{B} + \mathbf{E}$, where the columns of $\mathbf{A} \in \mathbb{R}^{d \times K}$ model the locations of each source (with $K$ sources total), the rows of $\mathbf{C} \in \mathbb{R}^{K \times T}$ model the time-varying fluorescence of each source, $\mathbf{B} \in \mathbb{R}^{d \times T}$ is a ``background" term to handle signals that can not easily be split into single-neuronal components, and $\mathbf{E} \in \mathbb{R}^{d \times T}$ denotes temporally and spatially uncorrelated noise. 

It is useful to break the processing pipeline into three sub-problems:
\begin{enumerate}
\item \textbf{Denoising}: separation of neural signal $\mathbf{Y}^{*} = \mathbf{A}\mathbf{C} + \mathbf{B}$ from noise $\mathbf{E}$;
\item \textbf{Compression} of signal $\mathbf{Y}^{*}$;  \item \textbf{Demixing}: factorization of $\mathbf{Y}^{*}$ into its constituent components $\mathbf{A},\mathbf{C}$, and $\mathbf{B}$.
\end{enumerate} 

Most prior work has attempted to solve these sub-problems simultaneously, e.g., to recover $\mathbf{A}$ and $\mathbf{C}$ directly from the raw data $\mathbf{Y}$.  As emphasized above, this direct approach involves a challenging non-convex optimization problem; the solution to this problem typically misses some structure in $\mathbf{Y}$, is highly sensitive to initialization and hyperparameter settings, and can be particularly unstable in low-SNR regimes.  We have found empirically that a sequential approach is more robust and effective.  First we compute the compressed and denoised estimate $\hat{\mathbf{Y}} = \mathbf{UV}$; here $\mathbf{U}$ and $\mathbf{V}$ are chosen so that $\hat{\mathbf{Y}}$ captures all of the signal in $\mathbf{Y}$ while retaining minimal noise (i.e.,  $\hat{\mathbf{Y}} \approx \mathbf{Y}^{*}$) and also $\mathbf{U}$ and $\mathbf{V}$ are highly-structured, compressible matrices, but we do not enforce any constraints between $(\mathbf{U}, \mathbf{V})$ and $(\mathbf{A}, \mathbf{C}, \mathbf{B})$.  The computation of $\mathbf{U}$ and $\mathbf{V}$ essentially solves sub-problems 1 and 2 simultaneously.  Second, we exploit $\mathbf{U}$, $\mathbf{V}$, and the resulting denoised $\hat{\mathbf{Y}}$ to facilitate the solution of problem 3.  We discuss each of these steps in turn below.



\subsection*{Denoising \& Compression}

To achieve good compression and denoising we need to take advantage of three key properties of functional imaging data:
\begin{enumerate}
\item Signal sources are (mostly) spatially local;
\item Signal is structured both temporally and spatially, whereas noise is temporally and spatially uncorrelated;
\item Signal is (mostly) low-rank.
\end{enumerate}
Given these structural assumptions, it is natural to construct $\mathbf{U}$ and $\mathbf{V}$ via a local penalized matrix decomposition approach\footnote{One important note: many matrix factorizations are possible here to obtain a compressed representation $(\mathbf{U},\mathbf{V})$.  This non-uniqueness does not pose an issue for either compression or denoising.  This makes these problems inherently easier than the demixing problem, where the identifiability of $\mathbf{A}$, $\mathbf{C}$, and $\mathbf{B}$ (perhaps up to permutations of the rows and columns of $\mathbf{A}$ and $\mathbf{C}$) is critical.}: we break the original data matrix $\mathbf{Y}$ into a collection of overlapping spatial patches, then decompose each of these matrix patches (in parallel) using a factorization method that enforces smoothness in the estimated spatial and temporal factors, then combine the resulting collection of spatial and temporal factors over all the patches into a final estimate of $\mathbf{U}$ and $\mathbf{V}$.  (See \href{https://github.com/flatironinstitute/CaImAn}{CaImAn} for a similar patch-wise approach to the demixing problem.)  


We have experimented with several approaches to penalized matrix decomposition (PMD), and found that an iterative rank-one deflation approach similar to the method described in \cite{Witten2009pmd} works well.  We begin by standardizing the data within a patch: for each pixel, we subtract the mean and normalize by an estimate of the noise variance within each pixel; the noise variance is estimated using the frequency-domain method described in \cite{pnevmatikakis2016simultaneous}, which exploits the fact that the signal and noise power are spectrally separated in movies with sufficiently high frame rates.  After this normalization we can model the noise $\mathbf{E}$ as roughly spatially and temporally homogeneous.  Denote this standardized data matrix within a patch as $\mathbf{Y_0}$, and Frobenius norm as $||.||_F$.  Then at the $k^{th}$ iteration PMD extracts the best rank-one approximation $\mathbf{u}_k\mathbf{v}_k^T$ to the current residual $\mathbf{R}_k = \mathbf{Y_0} - \sum_{n=1}^{k-1} \mathbf{u}_n\mathbf{v}_n^T$, as determined by the objective
\begin{equation}
(\mathbf{u}_k, \mathbf{v}_k) = \underset{\mathbf{u}, \mathbf{v}}{\arg\min} ~ || \mathbf{R}_k - \mathbf{u}  \mathbf{v}^T ||_F \hspace{1em} \text{subject to} \hspace{1em}
P_{spatial}(\mathbf{u}) \leq c_{1}^k,\  P_{temporal}(\mathbf{v}) \leq c_{2}^{k}, \label{eqn:CPMD}
\end{equation}
followed by a temporal debiasing update $\mathbf{v}_k = \mathbf{R}_k^T\mathbf{u}_k$.  The objective (\ref{eqn:CPMD}) can be ascended via alternating minimization on $\mathbf{u_k}$ and $\mathbf{v_k}$.


Note that if we drop the $P_{spatial}(\mathbf{u})$ and $P_{temporal}(\mathbf{v})$ constraints above then we can solve for $\mathbf{u}_k$ and $\mathbf{v}_k$ directly by computing the rank-1 singular value decomposition (SVD) of $\mathbf{R}_k$; in other words, by performing PCA within the patch.  Since we have normalized the noise scale within each pixel, PCA should identify the signal subspace within the patch, given enough data (because the normalized projected data variance in any direction will be equal to one plus the signal variance in this direction; since PCA searches for signal directions that maximize variance, PCA will choose exactly the signal subspace in the limit of infinite data).  Indeed, as discussed in the results section, simple patch-wise PCA (with an appropriate adaptive method for choosing the rank) often performs well, but incorporating spatial and temporal penalties in the optimization can push $\mathbf{u}_k$ and $\mathbf{v}_k$ closer to the signal subspace, resulting in improved compression and SNR.

How should we define the penalties $P_{spatial}(\mathbf{u})$ and $P_{temporal}(\mathbf{v})$, along with the corresponding constraints $c_{1}^k$ and $c_{2}^{k}$?  The simplest option would be to use quadratic smoothing penalties; this would lead to a simple closed-form linear smoothing update for each $\mathbf{u_k}$ and $\mathbf{v_k}$.  However, the signals of interest here have inhomogeneous smoothness levels --- an apical dendrite might be spatially smooth in the apical direction but highly non-smooth in the orthogonal direction, and similarly a calcium signal might be very smooth except at the times at which a spike occurs.  Therefore simple linear smoothing is typically highly suboptimal, often resulting in both undersmoothing and oversmoothing in different signal regions.  We have found total variation (TV) \cite{Rudin:1992:NTV:142273.142312} and trend filtering (TF) \cite{Kim2009tf} penalties to be much more empirically effective.  We let
\begin{align*}
P_{temporal}(\mathbf{v}) = \| \mathbf{D}^{(2)} \mathbf{v}\|_1 = \sum_{t=2}^{T-1} |\mathbf{v}_{t-1} - 2 \mathbf{v}_{t} + \mathbf{v}_{t+1}| 
\end{align*}
and
\begin{align*}
P_{spatial}(\mathbf{u}) = 
\|\mathbf{\nabla}_{\mathcal{G}}\mathbf{u}\|_1 = \sum_{(i,j) \in \mathcal{E}} | \mathbf{u}_i - \mathbf{u}_j |.
\end{align*}
Here $\mathbf{D}^{(2)}$ denotes the one-dimensional discrete second order difference operator and $\mathbf{\nabla}_{\mathcal{G}}$ the incidence matrix of the nearest-neighbor pixel-adjacency graph (pixels $(i,j)$ are in the edge set $\mathcal{E}$ if the pixels are nearest neighbors).

Similarly to \cite{pnevmatikakis2016simultaneous}, we define the smoothing constraints $c_1^k$ and $c_2^k$ implicitly within the alternating updates by the simple reformulation
\begin{equation}
\mathbf{u}_k = \underset{\mathbf{u}}{\arg\min} \| \mathbf{R}_k \mathbf{v}_k - \mathbf{u}\|_2^2\  s.t.\  \| \mathbf{\nabla}_{\mathcal{G}} \mathbf{u}\|_1 \leq c_1^k
\iff 
\mathbf{u}_k = \underset{\mathbf{u}}{\arg\min} \| \mathbf{\nabla}_{\mathcal{G}} \mathbf{u}\|_1\  s.t.\ \| \mathbf{R}_k \mathbf{v}_k - \mathbf{u}\|_2^2  \leq \hat{\sigma}^2_{\tilde{\mathbf{u}}} d \label{eqn:spatial_update}
\end{equation}
and 
\begin{equation}
\mathbf{v}_k = \underset{\mathbf{v}}{\arg \min} \| \mathbf{R}_k ^T \mathbf{u}_k - \mathbf{v}\|_2^2 \  s.t.\ \| \mathbf{D}^{(2)} \mathbf{v}\|_1 \leq c_2^k
\iff 
\mathbf{v}_k = \underset{\mathbf{v}}{\arg\min} \| \mathbf{D}^{(2)} \mathbf{v}\|_1 \ s.t.\ \| \mathbf{R}_k ^T \mathbf{u}_k - \mathbf{v}\|_2^2  \leq \hat{\sigma}^2_{\tilde{\mathbf{v}}} T \label{eqn:temporal_update}
\end{equation}
where $\hat{\sigma}^2_{\tilde{\mathbf{u}}}$ (resp.~$\hat{\sigma}^2_{\tilde{\mathbf{v}}}$) estimates the noise level of the unregularized update $\tilde{\mathbf{u}}_k = \mathbf{R}_k \mathbf{v}_k$ (resp.~$\tilde{\mathbf{v}}_k = \mathbf{R}_k^T \mathbf{u}_k$), and we are using the fact that if the residual $\mathbf{R}_k \mathbf{v}_k - \mathbf{u}$ contains just noise then its squared norm should be close to $\hat{\sigma}^2_{\tilde{\mathbf{u}}} d$, by the law of large numbers (and similarly for equation \ref{eqn:temporal_update}).   See Algorithm \ref{alg:ROD} for a summary.

To solve the constrained problems on the right-hand side we use the line search approach described in \cite{Langer2017cps}.  We solve the primal form of the TV optimization problem (\ref{eqn:spatial_update}) using the proxTV package \cite{barberoTV14}, and of the TF optimization problem (\ref{eqn:temporal_update}) using the Primal-Dual Active Set method in \cite{Han2016pdas}.  Both of these methods can exploit warm starts, leading to major speedups after a good initial estimate is found.  Empirically the TF optimization scales linearly with the movie length $T$; since the scale of the TV problem is bounded (because we work in local spatial patches) we have not explored the scaling of the TV problem in depth.

\RestyleAlgo{boxruled}
\LinesNumbered
\begin{algorithm}[t!] 
	\caption{Pseudocode for performing Single Factor PMD(TV,TF) (\ref{eqn:CPMD}).} 
	\label{alg:ROD} 
    \Fn{Rank One Approximation$(\mathbf{R} \in \mathbb{R}^{d\times T}):$}{
      \begin{algorithmic}[1] 
          \STATE $\mathbf{u}_0 \leftarrow Decimated\ Initialization(\mathbf{R})$;
          \STATE $\mathbf{v}_0\leftarrow Temporal\ Update (\mathbf{R}, \mathbf{u}_0)$;
          \STATE $n \leftarrow 0$;
          \WHILE{$\min(\|\mathbf{u}_{n} - \mathbf{u}_{n-1}\|_2,\ \|\mathbf{v}_{n} - \mathbf{v}_{n-1}\|_2) >$ tol}
          \STATE $\mathbf{u}_{n+1} \leftarrow Spatial\ Update (\mathbf{R}, \mathbf{v}_{n})$;
          \STATE $\mathbf{v}_{n+1} \leftarrow Temporal\ Update (\mathbf{R}, \mathbf{u}_{n+1})$;
          \STATE $n \leftarrow n + 1$;
          \ENDWHILE
      \end{algorithmic}
      \Sr{Decimated Initialization$(\mathbf{R}  \in \mathbb{R}^{d\times T}):$}{
      \begin{algorithmic}[1] 
      	  \STATE $\mathbf{R}_{ds} \leftarrow Decimate(\mathbf{R})$
		  \STATE $\mathbf{u}_0 \leftarrow \mathbf{1} / \|\mathbf{1}\|_2$;
          \STATE $\mathbf{v}_0 \leftarrow \mathbf{R}_{ds}^T \mathbf{u}_0 / \|\mathbf{R}_{ds}^T \mathbf{u}_0\|_2$;
          \STATE $n \leftarrow 0$;
          \WHILE{$\min(\|\mathbf{u}_{n} - \mathbf{u}_{n-1}\|_2,\ \|\mathbf{v}_{n} - \mathbf{v}_{n-1}\|_2) >$ tol}
          \STATE $\mathbf{u}_{n+1} \leftarrow \mathbf{R}_{ds} \mathbf{v}_{n} / \|\mathbf{R}_{ds} \mathbf{v}_{n}\|_2$;
          \STATE $\mathbf{v}_{n+1} \leftarrow  \mathbf{R}_{ds}^T \mathbf{u}_{n+1} / \|\mathbf{R}_{ds}^T \mathbf{u}_{n+1}\|_2$;
          \STATE $n \leftarrow n + 1$;
          \ENDWHILE
          \STATE $\mathbf{u}_n \leftarrow Upsample(\mathbf{u}_n)$
          \RETURN $\mathbf{u}_n$
      \end{algorithmic}
      }
      \Sr{Spatial Update$(\mathbf{R}  \in \mathbb{R}^{d\times T}, \mathbf{v} \in \mathbb{R}^{T}):$}{
      \begin{algorithmic}[1] 
          \STATE $\tilde{\mathbf{u}} \leftarrow \mathbf{R} \mathbf{v}$;
          \STATE $\hat{\sigma}^2_{\mathbf{u}} \leftarrow$ Image Noise Estimate$(\tilde{\mathbf{u}})$;
          \STATE $\mathbf{u}\leftarrow \arg\min_\mathbf{u} \| \mathbf{\nabla}_{\mathcal{G}}\mathbf{u}\|_{1} \ s.t. \ \| \tilde{\mathbf{u}} - \mathbf{u} \|_2^2 \leq \hat{\sigma}^2_{\mathbf{u}} d$;
          \RETURN $\mathbf{u} / \|\mathbf{u}\|_2$
      \end{algorithmic}
      }
      \Sr{Temporal Update$(\mathbf{R} \in \mathbb{R}^{d\times T}$,  $\mathbf{u} \in \mathbb{R}^{d}):$}{
      \begin{algorithmic}[1] 
          \STATE $\tilde{\mathbf{v}}\leftarrow \mathbf{R}^T \mathbf{u}$;
          \STATE $\hat{\sigma}^2_{\mathbf{v}} \leftarrow$Timeseries Noise Estimate$(\tilde{\mathbf{v}})$;
          \STATE $\mathbf{v}\leftarrow \arg\min_\mathbf{v} \| \mathbf{D}^{(2)}\mathbf{v}\|_{1} \ s.t. \ \| \tilde{\mathbf{v}} - \mathbf{v} \|_2^2 \leq \hat{\sigma}^2_{\mathbf{v}}T$;
          \RETURN $\mathbf{v} / \|\mathbf{v}\|_2$
      \end{algorithmic}
    }
    }
\end{algorithm}

\begin{figure}[t!]
\centering
	\includegraphics[width=17cm]{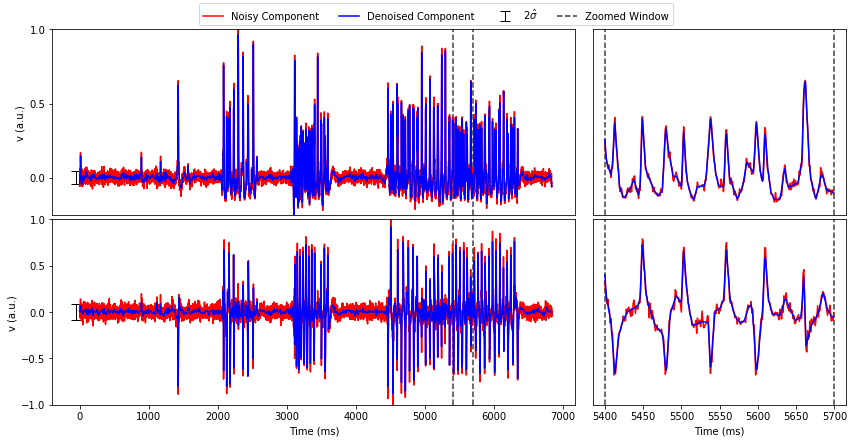}
	\caption{Illustration of trend filtering.  Each row shows a component $\mathbf{v}$ extracted from the voltage imaging dataset (see Results section for details).  Red indicates simple projected signal $\tilde{\mathbf{v}} = \mathbf{R}^T \mathbf{u}$; blue indicates $\mathbf{v}$ after trend filtering.  Errorbars on left indicate $2 \times$ estimated noise scale; right panels show zoomed region indicated by dashed lines in left panel.}
    \label{fig:TrendFiltering}
\end{figure}

Figure \ref{fig:TrendFiltering} illustrates the effect of trend filtering on a couple $\mathbf{v}$ components.  One important difference compared to previous denoising approaches \cite{haeffele2014structured,pnevmatikakis2016simultaneous} is that the TF model is more flexible than the sparse autoregressive model that is typically used to denoise calcium imaging data: the TF model does not require the estimation of any sparsity penalties or autoregressive coefficients, and  can handle a mixture of positive and negative fluctuations, while the sparse nonnegative autoregressive model can not (by construction).  This is  important in this context since each component in $\mathbf{V}$ can include multiple cellular components (potentially with different timescales), mixed with both negative and positive weights.

To complete the description of the algorithm on a single patch we need an initialization and a stopping criterion to adaptively choose the rank of $\mathbf{U}$ and $\mathbf{V}$.  For the latter, the basic idea is that we want to stop adding components $k$ as soon as the residual looks like uncorrelated noise.  To make this precise, we define a pair of spatial and temporal ``roughness" test statistics
\begin{align*}
&T_{temporal}(\mathbf{v}) = \|\mathbf{D}^{(2)} \mathbf{v}\|_1 / \| \mathbf{v} \|_1
&T_{spatial}(\mathbf{u}) = \|\mathbf{\nabla}_{\mathcal{G}}\mathbf{u}\|_1 / \| \mathbf{u} \|_1
\end{align*}
and compute these statistics on each extracted $\mathbf{u}_k$ and $\mathbf{v}_k$.  We accept or reject each component according to a one-sided hypothesis test under the null hypothesis that $\mathbf{R}_k$ consists of uncorrelated Gaussian noise of variance one.  (We compute the critical region for this test numerically.)  In the compression stage we are aiming to be rather conservative (we are willing to accept a bit of extra noise or a slightly higher-rank $\mathbf{U}$ and $\mathbf{V}$ in order to ensure that we are capturing the large majority of the signal), so we terminate the outer loop (i.e., stop adding more components $k$) after we reject a couple components $k$ in a row.  See Algorithm \ref{alg-full-PMD} for a summary.

To initialize, we have found that setting $\mathbf{u}_0 \propto \mathbf{1}$ works well.  To speed up early iterations, it is natural to iterate the projections while skipping the denoising steps; this corresponds to intializing with an approximate rank-1 SVD as computed by power iterations. Initializing in this manner can reduce the total number of iterations needed for $\mathbf{u}_k, \mathbf{v}_k$ to converge.  Matrix-vector multiplications are a rate limiting step here; thus, these initial iterations can be sped up using spatial and temporal decimation on $\mathbf{R}_k$.  Empirically, decimation has the added benefit of boosting signal (by averaging out noise in neighboring timepoints and pixels) and can be useful for extracting weak components in low SNR regimes; see \cite{friedrich2017multi} for a related discussion.



\begin{algorithm}[t!]
	\caption{Pseudocode for Full PMD(TF,TV).}
    \label{alg-full-PMD}
    \Fn{Compress Patch$(Y\in \mathbb{R}^{d \times T}, spatial\_thresh, temporal\_thresh):$}{
		\begin{algorithmic}[1] 
          \STATE $\mathbf{U} \leftarrow [\ ],\ \mathbf{V} \leftarrow [\ ],\ \mathbf{R} \leftarrow \mathbf{Y}$;
          \STATE $num\_fails\leftarrow 0$;
          \WHILE{$num\_fails< max\_num\_fails$}
          \STATE $\mathbf{u}, \mathbf{v} \leftarrow Rank\ One\ Decomposition(\mathbf{R})$;
          \STATE $\mathbf{v} \leftarrow \mathbf{R}^T \mathbf{u}$; \hspace{2em}// debias \& rescale
          \IF {$\|\mathbf{\nabla}_{\mathcal{G}}\mathbf{u}\|_{1} / \|\mathbf{u}\|_1 < spatial\_thresh $ and $\|\mathbf{D}^{(2)}\mathbf{v}\|_{1} / \|\mathbf{v}\|_1 < temporal\_thresh$}
          \STATE $\mathbf{U} \leftarrow [\mathbf{U}, \mathbf{u}], \mathbf{V}\leftarrow [\mathbf{V}, \mathbf{v}]$,\ $num\_fails \leftarrow 0$;
          \ELSE
          \STATE $num\_fails\leftarrow num\_fails + 1$
          \ENDIF
          \STATE $\mathbf{R} \leftarrow \mathbf{R} - \mathbf{u} \mathbf{v}^T$;
          \ENDWHILE
          \RETURN $\mathbf{U}, \mathbf{V}$
		\end{algorithmic}
    }
\end{algorithm}

The method described so far handles a single spatial patch of data.  We can process patches in parallel; a multi-core implementation of this method (assigning different patches to different cores) achieves nearly linear speedups.  We have found that for some datasets edge artifacts can appear near patch boundaries if the patches do not overlap spatially.  These boundary artifacts can be eliminated by performing a $4 \times$ over-complete block-wise decomposition of $\mathbf{Y}$ using half-offset grids for the partitions (so that each pixel $x$  lies within the interior of at least one patch). Then we combine the overlapping patches together via linear interpolation (see \cite{pnevmatikakis2017normcorre} for a similar approach): set 
\begin{equation*}
\hat{\mathbf{Y}}(x,t) = \frac{\sum_{p} \mathbf{a}_p(x) \hat{\mathbf{Y}}_p(x,t)} {\sum_p \mathbf{a}_p(x)}, 
\end{equation*}
where $p$ indexes the patches (so $\hat{\mathbf{Y}}_p$ denotes the denoiser output in the $p$-th patch) and $0 \leq \mathbf{a}_p(x) \leq 1$ is a ``pyramid" function composed of piecewise linear functions that start at $0$ at the patch boundaries and increase linearly to $1$ at the center of the patch. 

The above is equivalent to starting with a collection of overlapping sparse local factorizations $\mathbf{U}_p \mathbf{V}_p$, forming element-wise products between the individual spatial components $\mathbf{U}_{ip}$ and the pyramid functions $\mathbf{a}_p$, and then forming the union of the result to obtain a new factorization $\mathbf{UV}$. Typically this will result in some redundancy due to the overlapping spatial components; we remove this redundancy in a final backwards model selection step that tests whether each temporal component can be explained as a weighted sum of its neighbors. More precisely, we sort the components in ascending order according to the $L_2$ norms of $\mathbf{U}_{ip} \cdot a_p$.  For each $i$ in this order we then regress $\mathbf{V}_i$ onto the collection of temporal components $\mathbf{V}_j$ whose corresponding spatial components $\mathbf{U}_j$ overlap with $\mathbf{U}_i$, i.e., approximate $ \hat{\mathbf{V}_i} = \sum_j \beta_j \mathbf{V}_j$. We then test the signal strength of the residual $\mathbf{V}_i - \hat{\mathbf{V}_i}$ (using the temporal test statistic defined previously); 
the component is rejected if the residual is indistinguishable from noise according to this test statistic.  If component $i$ is rejected then we  distribute its energy to the remaining spatial components according to the regression weights: $\mathbf{U}_{j} = \mathbf{U}_{j} + \beta_{j} \mathbf{U}_{i}$.

We conclude with a few final implementation notes.  First, the results do not depend strongly on the precise patch size, as long as the patch size is comparable to the spatial correlation scale of the data: if the patches are chosen to be much smaller than this then the $\mathbf{V}$ components in neighboring patches are highly correlated, leading to excessive redundancy and suboptimal compression.  (Conversely, if the patch size is too big then the sparsity of $\mathbf{U}$ is reduced, and we lose the benefits of patch-wise processing.)

Second, in some datasets (e.g., widefield imaging, or microendoscopic imaging data), large background signals are present across large portions of the field of view.  These background signals can be highly correlated across multiple spatial patches, leading to a suboptimal compression of the data if we use the simple independent-patch approach detailed above.  Thus in some cases it is preferable to run a couple iterations of PMD(TV, TF) on the full $\mathbf{Y}$ and then subtract the resulting components away before moving on to the independent block processing scheme.  We have found that this effectively subtracts away dominant background signals; these can then be encoded as a small number of dense columns in the matrix $\mathbf{U}$, to be followed by a larger number of sparse columns (corresponding to the small patches), resulting in an overall improvement in the compression rate.  See the \hyperref{\VideoEndoscopeBkgURL}{}{pmd_videoendoscopebkg}{microendoscopic imaging background video} for an example.

The patch-wise PMD(TV,TF) approach results in an algorithm that scales linearly in three critical parameters: $T$ (due to the sparse nature of the second-difference operator in the TF step), $d$ (due to the patch-wise approach), and the rank of $\mathbf{U}$ and $\mathbf{V}$.  We obtain further speedups by exploiting warm starts and parallel processing over patches.  Additional speedups can be obtained for very long datasets by computing $\mathbf{U}$ on a subset of the data and then updating $\mathbf{V}$ on the remainder of the movie; the latter step does not require any PMD iterations (since the spatial signal subspace has already been identified) and is therefore very fast, just requiring a single temporal update call per element of $\mathbf{V}$.

\subsection*{Demixing}

The methods described above provide a compressed and denoised representation of the original data $\mathbf{Y}$: the output matrices $\mathbf{U}$ and $\mathbf{V}$ are low-rank compared to $\mathbf{Y}$, and $\mathbf{U}$ is additionally highly sparse (since $\mathbf{U}$ is formed by appending spatial components $\mathbf{u}$ from multiple local spatial patches, and each $\mathbf{u}_k$ is zero outside of its corresponding patch).  How can we exploit this representation to improve the demixing step?

It is useful to first take a step back to consider the strengths and weaknesses of current state of the art demixing methods, most of which are based on NMF.  The NMF model is very natural in calcium imaging applications, since each neuron has a shape that is fixed over the timescale of a typical imaging experiment (and these shapes can be represented as non-negative images, i.e., an element of the $\mathbf{A}$ matrix), and a corresponding time-varying calcium concentration that can be represented as a non-negative vector (an element of $\mathbf{C}$): to form a movie we simply take a product of each of these terms and add them together with noise and background, i.e., form $\mathbf{Y}= \mathbf{AC} + \mathbf{B} + \mathbf{E}$.


However, current NMF-based approaches leave room for improvement in several key directions.  First, since NMF is a non-convex problem, good initializations are critical to obtain good results via the standard alternating optimization approaches (similar points are made in \cite{petersen2017scalpel}).  Good initialization approaches have been developed for somatic or nuclear calcium imaging, where simple Gaussian shape models are useful crude approximations to the elements of $\mathbf{A}$ \cite{pnevmatikakis2016simultaneous}, but these approaches do not apply to dendritic or axonal imaging.  Second (related), it can be hard to separate weak components from noise using current NMF-based approaches.  Finally, voltage imaging data does not neatly fit in the NMF framework, since voltage traces typically display both positive and negative fluctuations around the baseline resting potential.

To improve the robustness of NMF approaches for demixing functional data, we make use of the growing literature on ``guaranteed NMF'' approaches --- methods for computing a non-negative matrix factorization that are guaranteed to output the ``correct'' answer under suitable conditions and assumptions \cite{donoho2004does,recht2012factoring,arora2012computing,li2016recovery}. In practice, these methods work well on clean data of sufficiently small dimensionality, but are not robust to noise and scale poorly to high-dimensional data.  We can solve both of these issues by ``superpixelizing" the denoised version of $\mathbf{Y}$; the resulting NMF initialization method improves significantly on state of the art methods for processing dendritic and axonal data.  We also take advantage of the sparse, low-rank structure of $\mathbf{U}$ and $\mathbf{V}$ to speed up the NMF iterations.


\subsubsection*{Initialization via pure superpixels}

  
The first step of the initialization procedure is to identify groups of highly correlated spatially connected pixels -- ``superpixels."  The idea is that a pixel within a neuron should be highly correlated with its neighbors, while a pixel containing mostly noise should have a much lower neighbor correlation.  These neighbor correlations, in turn, can be estimated much more accurately from the denoised compared to the raw data.  The superpixelization procedure results in a set of non-overlapping groups of pixels which are likely to be contained in good neural components. Then we want to extract ``pure'' superpixels, i.e., the subset of superpixels dominated by signal from just one neural component.  We will use the temporal signals extracted from these pure superpixels to seed $\mathbf{C}$ in the NMF decomposition.

To identify superpixels, we begin with the denoised data $\hat{\mathbf{Y}} = \mathbf{UV}$.  Since the compression process discussed in the previous section is rather conservative (aiming to preserve the full signal, at the expense of retaining a modest amount of noise), there is room to apply a more aggressive lossy denoiser in the initialization stage to further reduce any remaining noise in $\hat{\mathbf{Y}}$.  We soft-threshold signals in each pixel that are not sufficiently large --- less than the median plus $\delta \times$ the median absolute deviation (MAD) within each pixel, with $\delta \approx 1$ or $2$.  (This thresholding serves to extract mostly spiking activity from functional imaging data.)  We identify two neighboring pixels to be from the same superpixel if their resulting denoised, soft-thresholded temporal signals have a correlation larger than a threshold $\epsilon$, with $\epsilon \approx 0.9$.  Superpixels that contain fewer than $\tau$ pixels are discarded to further reduce noise and the total number of superpixels.  We then apply rank 1 NMF on the signals from each superpixel to extract their (thresholded) temporal activities.

To extract pure superpixels, we apply the Successive Projection Algorithm (SPA) \cite{gillis2014fast} to the temporal activities of superpixels.  This algorithm removes ``mixed'' superpixels whose temporal activity can be modeled as a nonnegative linear combination of activity in other superpixels (up to some R-squared level larger than $ 1-\kappa$, where we use $\kappa \approx 0.2$) and outputs the remaining ``pure" superpixels.  See Algorithm \ref{alg1} for pseudocode.

\begin{figure}[t!]
	\centering
	\begin{tabular}{rc}
	\ctab{A} & 	\includegraphics[width=1\textwidth,clip = true, trim = 5.7cm 0.55cm 0.23cm 0.6cm]{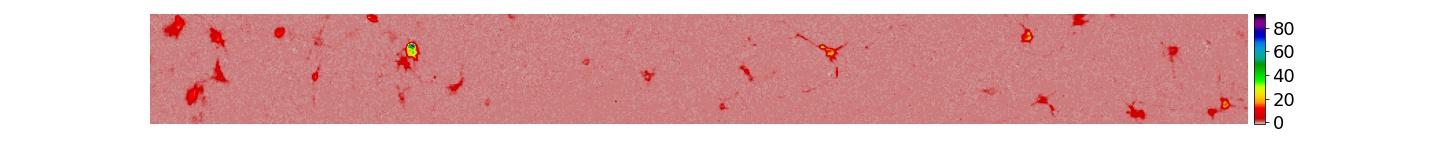}\\
	\ctab{B} & \includegraphics[width=1\textwidth,clip = true, trim = 5.5cm 0.5cm 0.2cm 0.67cm]{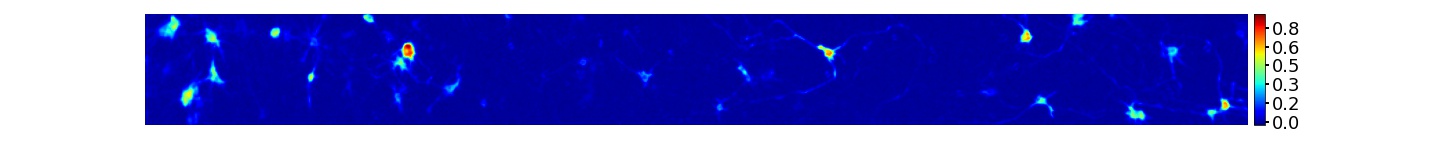}\\
		\ctab{C} & \includegraphics[width=1\textwidth,clip = true, trim = 5cm 0.5cm -1.6cm 0.6cm]{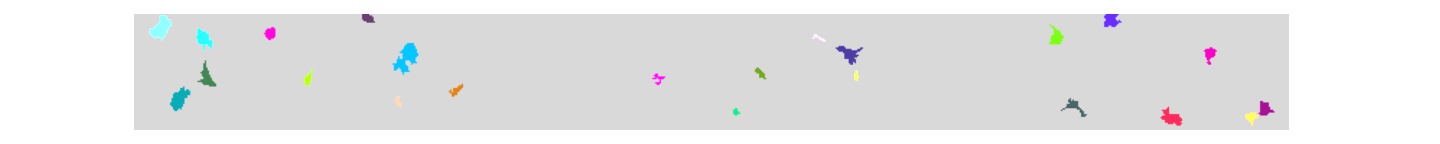}\\
\ctab{D} & \includegraphics[width=1\textwidth,clip = true, trim = 5.7cm 0.54cm 0.5cm 0.6cm]{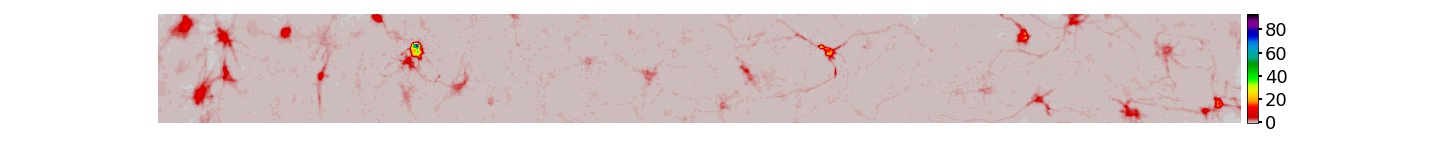}\\
\ctab{E} & \includegraphics[width=1\textwidth,clip = true, trim = 4.2cm 0.5cm -1cm 0.65cm]{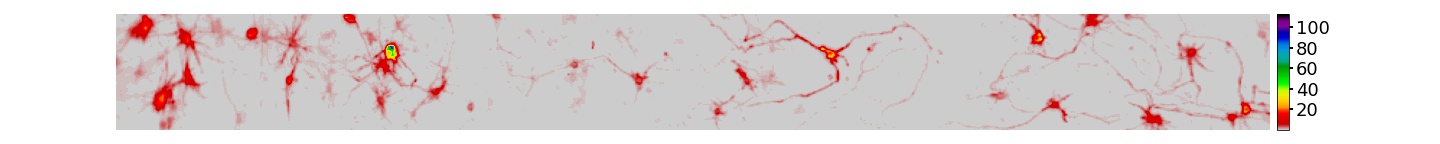}\\
\ctab{F} & \includegraphics[width=1\textwidth,clip = true, trim = 4.9cm 0.5cm -0.4cm 0.6cm]{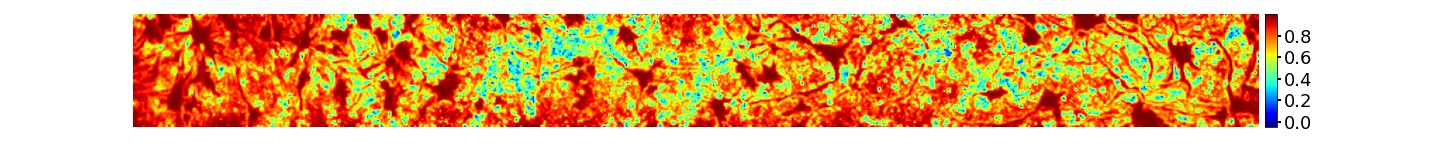}\\	
		\ctab{G} & \includegraphics[width=1\textwidth,clip = true, trim = 5cm 0.5cm -1.5cm 0.6cm]{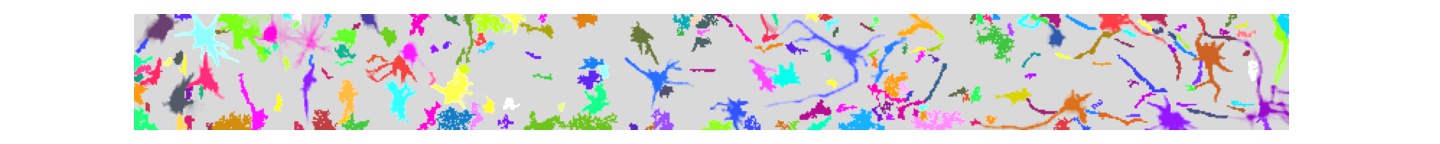}\\
		\ctab{H} & \includegraphics[width=1\textwidth,clip = true, trim = 5cm 0.5cm -1.5cm 0.6cm]{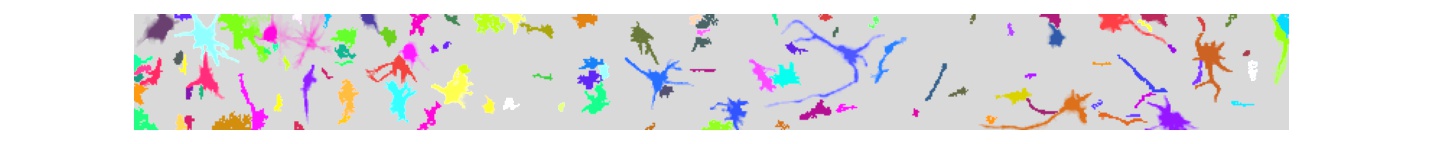}\\
	\end{tabular}
\caption{Denoising helps extract more complete superpixels in voltage imaging data (see Appendix for full dataset details). (A) Mean intensity projection of detrended data $\mathbf{Y}$.  (A spline detrender was applied to the raw data prior to analysis; see Appendix for details. This detrending should not be confused with an application of the trend filtering denoiser.) (B) Local correlation image of detrended data $\mathbf{Y}$. (C) Superpixels extracted in detrended data $\mathbf{Y}$ with correlation cut-off $\epsilon=0.2$, size cut-off $\tau=10$. (D) Mean intensity projection of denoised data $\hat{\mathbf{Y}}$. (E) Mean intensity projection of soft-thresholded denoised data. (F) Local correlation image of soft-thresholded denoised data; note that neural shapes are much clearer here than in panel A. (G) Superpixels extracted in soft-thresholded data with correlation cut-off $\epsilon=0.95$, size cut-off $\tau=15$.  Note that we are using much more stringent criteria for defining superpixels here compared to panel C, but nonetheless (due to denoising) extract a much more complete superpixelization. (H) ``Pure" superpixels extracted in soft-thresholded data with $\tau=0.2$.  See the \href{\VideoSuperpixelizationURL}{superpixelization video} for a time-varying illustration of these processing steps.}
\label{fig:vi_superpixels}
\end{figure}

Note that running SPA on superpixels rather than raw pixels improves performance significantly here, since averaging signals within superpixels boosts SNR (making it easier to separate signal from noise and isolate pure from mixed pixels) and also greatly reduces the dimensionality of the non-negative regression problem SPA has to solve at each iteration.  (To keep the problem size small we also run SPA just on small local spatial patches, as in the previous section.)  Finally, while we have obtained good results with SPA, other approaches are available \cite{gillis2018fast} and could be worth further exploration in the future.  See Figure \ref{fig:vi_superpixels} for a visual summary of the full procedure. 

\subsubsection*{Local NMF}
Next we run NMF, using the temporal signals extracted from the ``pure'' superpixels to initialize $\mathbf{C}$.  Given the initial $\mathbf{C}$, the typical next step is to regress onto the data to initialize $\mathbf{A}$.  (Note that pure superpixels typically capture just a subset of pixels within the corresponding neuron, so it is not efficient to initialize $\mathbf{A}$ with the pure superpixels.)  However, given the large number of pixels in a typical functional imaging video, direct regression of $\mathbf{C}$ onto $\mathbf{Y}$ is slow and overfits, providing poor estimates of $\mathbf{A}$.

This issue is well-understood \cite{pnevmatikakis2016simultaneous}, and several potential solutions have been proposed.
For somatic imaging it makes sense to restrict the support of $\mathbf{A}$ to remain close to their initial values (we could use a dilation of the superpixel support for this).  But for data with large dendritic or axonal components this approach would cut off large fractions of these components.  Sparse regression updates are an option here, but these do not enforce spatial structure in the resulting $\mathbf{A}$ directly; this often results in ``speckle" noise in the estimated spatial components (c.f.\ Figure \ref{realcompare} below).

We have found the following approach to be more effective.  We initialize the support set $\Omega_k$ as the support of the $k$-th ``pure'' superpixel.  Given $\mathbf{C}$, we compute the correlation image for each component $k$ as the correlation between the denoised data $\hat{\mathbf{Y}}$ and the $k$-th temporal component, $\mathbf{C}_k$.  We truncate this correlation image below a certain threshold $\epsilon_1$ to zero, then update $\Omega_k$ as the connected component of the truncated correlation image which overlaps spatially with the previous $\Omega_k$.  We use the modified fastHALS algorithm in \cite{friedrich2017multi} to update $\mathbf{A}$, $\mathbf{C}$, and $\mathbf{B}$ to locally optimize the objective
\begin{equation}\label{objnmf}
	\min_{\mathbf{A},\mathbf{C},\b} \|\hat{\mathbf{Y}}-\mathbf{AC} -\mathbf{B}\|_{F}^2,\ \mathrm{s.t.}\ \mathbf{A}_k^x = 0 ~ \forall x \not \in \Omega_k , \mathbf{A}\geqslant 0, \mathbf{C}\geqslant 0, \mathbf{B}=\b\mathbf{1}^T, \b\geqslant 0.
\end{equation}
Here we have modeled the background $\mathbf{B}$ as a simple temporally-constant vector; we discuss generalizations to time-varying backgrounds below.  Also note that we are approximating $\hat{\mathbf{Y}}$ directly here, not the thresholded version we used to extract the superpixels above.

Finally, we incorporate a merge step: we truncate the correlation image below certain threshold $\epsilon_2$ to zero, and automatically merge neurons if their truncated correlation images are highly overlapped.  The full algorithm is shown in Algorithm \ref{alg2}. 


\subsubsection*{Further implementation details}
\textit{Multi pass strategy:} As in \cite{zhou2018efficient}, we find it effective to take a couple passes over the data; particularly in datasets with high neuron density, the first NMF pass might miss some dim neurons.  We decrease the MAD threshold $\delta$ and re-run Algorithm \ref{alg1} on the residual to find additional components, and then run a final merge and NMF update to complete the pipeline.

\textit{Improvements from denoising and compression:}  Compressed data leads to faster NMF updates, since we can replace $\hat{\mathbf{Y}}$ as $\mathbf{UV}$; in fastHALS, we can regress each $\a_k$ on $\mathbf{U}$ or $\c_{k}$ on $\mathbf{V}$ first instead of directly onto $\mathbf{Y}$.  Similarly, when calculating the correlation image, we can compute the correlation between the low rank $\mathbf{V}$ and $\c_k$ first.  As emphasized above, denoising also improves the estimation of the correlation images, which in turn improves the estimation of the support sets $\Omega_k$.

\textit{Time-varying background:} It is straightforward to generalize the objective \ref{objnmf} to include a time-varying background, using  either a low-rank model (as in \cite{pnevmatikakis2016simultaneous}) or a ring-structured model (as in \cite{zhou2018efficient}).  For the low-rank background model, we have found that performing an SVD on the data excluding the support of the superpixels provides an efficient initialization for the background temporal components.



\textit{Incorporating temporal penalties}: Note that we are only imposing nonnegativity in $\mathbf{C}$ here; after denoising to obtain $\hat{\mathbf{Y}}$, we have found that this simple nonnegative constraint is sufficient for the datasets examined here.  However, it is certainly possible to incorporate temporal penalties or constraints on $\mathbf{C}$ (e.g., a TF penalty or a non-negative auto-regressive penalty as in \cite{pnevmatikakis2016simultaneous}), either within each iteration or as a final denoising step.

\textit{Post-processing}: We find that sorting the extracted components by their ``brightness," computed as $\max \a_k\cdot\max\c_k$, serves to separate dim background components from bright single-neuronal components.  
We also found it useful to drop components whose temporal trace has skewness less than 0.5; traces with high skewness correspond to components with significant spiking activity, but low-skewness traces corresponded to noise.  

\begin{algorithm}[t!]
	\caption{Pseudocode for the complete proposed pipeline.} 
	\label{alg1} 
	\begin{algorithmic}[1] 
		\REQUIRE Motion corrected data $\mathbf{Y}\in \mathbb{R}^{d\times T}$, MAD threshold $\delta$, minimum size of superpixels $\tau$, correlation threshold for superpixels $\epsilon$,  $R^2$ threshold in SPA $\kappa$. 
		\STATE $\sigma(\x)\leftarrow$ estimated noise for each pixel $\x$ of $\mathbf{Y}$;
		\STATE $\mu(\x)\leftarrow$ mean for each pixel of $\mathbf{Y}$; 
		\STATE $\mathbf{Y} \leftarrow \left(\mathbf{Y}-\mu(\x)\right) / \sigma(\x)$;
		\STATE $(\hat{\mathbf{Y}},\mathbf{U},\mathbf{V}) \leftarrow$ PMD($\mathbf{Y}$);
		\STATE $n \leftarrow 0$; $\mathbf{A} \leftarrow [\ ]$, $\mathbf{C}\leftarrow [\ ]$, $\b\leftarrow\mathrm{median}$ for each pixel of $\hat{\mathbf{Y}}$;
		\WHILE{$n < $ maximum number of passes}
		\STATE $\mathbf{R} \leftarrow \hat{\mathbf{Y}} -\mathbf{AC} - \b$;
		\STATE $\sigma_{med}(\x)\leftarrow$ median absolute deviation for each pixel of $\mathbf{R}$;
		\STATE $\mu_{med}(\x)\leftarrow$ median for each pixel of $\mathbf{R}$;
		\STATE $\tilde{\mathbf{Y}}\leftarrow \max\left(0, \mathbf{R} - \mu_{med}(\x) - \delta\cdot\sigma_{med}(\x)\right)$;
		\STATE $\mathrm{corr}(\x,\x^*)\leftarrow \mathrm{corr}\left(\tilde{\mathbf{Y}}(\x,t),\tilde{\mathbf{Y}}(\x^*,t)\right)$ for all neighbouring pixel pairs $(\x,\x^*)$;
		\STATE Extract superpixels: connect $\x$ and $\x^*$ together if $\mathrm{corr}(\x,\x^*)\geqslant \epsilon$ to construct connected components and discard those smaller than $\tau$, forming superpixels $\Omega_k,k=1,\cdots,K$;
		\STATE $(\a _{k}, \c_{k})\leftarrow \mathrm{rank\ 1\ NMF}$ of $\tilde{\mathbf{Y}}$ on support $\Omega_k , k= 1,\cdots, K$;
		\STATE $[i_1,i_2,\cdots,i_S]\leftarrow\mathrm{SPA}([\c_{1},\c_{2},\cdots,\c_{K}], \kappa)$;  $i_1,i_2,\cdots,i_S$ are indices of pure superpixels;
		\STATE $\mathbf{A}_0\leftarrow[\mathbf{A}, \a_{i_1},\a_{i_2},\cdots,\a_{i_S}]$;
		\STATE $\mathbf{C}_0\leftarrow[\mathbf{C}^T, \c_{i_1},\c_{i_2},\cdots,\c_{i_S}]^T$;
        \STATE $\b_0\leftarrow \b$;
		\STATE $(\mathbf{A}, \mathbf{C}, \b)\leftarrow\mathrm{LocalNMF}(\mathbf{U}, \mathbf{V}, \mathbf{A}_0, \mathbf{C}_0, \b_0)$;
		\STATE $\delta \leftarrow \delta-1$;
		\STATE $n \leftarrow n+1$;		
		\ENDWHILE
        \STATE $\eta(k)\leftarrow$ estimated noise for $\c_k$ using average of high frequency domain of PSD;
        \STATE (Optional) Denoise temporal components, e.g.~by $\ell_1$ trend filter: $\c_k\leftarrow \min\limits_{\tilde{\c}_k} \|\tilde{\c}_k\|_1,\ \mathrm{s.t.}\ \|\tilde{\c}_k-\c_k\|_{F}\leqslant \eta(k)\sqrt{T}, k=1,\cdots,K$;
		\RETURN $\mathbf{A},\mathbf{C},\b$
	\end{algorithmic}
\end{algorithm}

\begin{algorithm}
	\caption{Pseudocode for LocalNMF.}
	\label{alg2}
	\begin{algorithmic}[1]
		\REQUIRE Compressed factors $\mathbf{U} \in \mathbb{R}^{d\times r}, \mathbf{V} \in \mathbb{R}^{T\times r}$ ($r = rank (\hat{\mathbf{Y}})$); initial constant background $\b_0$, spatial components $\mathbf{A}_0=[\a_{1,0},\cdots,\a_{K,0}]\in\mathbb{R}^{d\times K}$, and temporal components $\mathbf{C}_0=[\c_{1,0},\cdots,\c_{K,0}]^T \in\mathbb{R}^{K\times T}$; truncation threshold when updating support $\epsilon_1$, truncation threshold when merging $\epsilon_2$, overlap threshold when merging $\epsilon_3$.
		\STATE $\Omega_k \leftarrow \mathrm{supp}(\a_{k,0})$ is spatial support for $k$-th component, $k=1,\cdots,K$; 
		\STATE $\hat{\mathbf{A}} \leftarrow \mathbf{A}_0, \hat{\mathbf{C}}\leftarrow \mathbf{C}_0, \hat{\b}\leftarrow\b_0$;
        \STATE $\nu(\x)\leftarrow$ standard deviation for each pixel of $\hat{\mathbf{Y}} = \mathbf{UV}$; 
        \STATE $\bar{\mathbf{V}}\leftarrow$ mean for each column of $\mathbf{V}$;        
		\WHILE{not converged}
        \STATE$\mathbf{P} \leftarrow \left[\mathbf{U},-\b\right]\left(
        \begin{bmatrix}
        \mathbf{V}\\
        \mathbf{1}^T\\
		\end{bmatrix}\hat{\mathbf{C}}^T\right)$;
        \STATE$\mathbf{Q} \leftarrow \hat{\mathbf{C}}\hat{\mathbf{C}}^{T}$;
		\FOR{$k=1:K$}
		\STATE Update spatial: $\hat{\a}_{k}(\Omega_k) \leftarrow \max\left(0, \hat{\a}_{k}(\Omega_k) + \frac{\mathbf{P}(\Omega_k,k)-\hat{\mathbf{A}}(\Omega_k)\mathbf{Q}(:,k)}{\mathbf{Q}(k,k)}\right)$;
		\ENDFOR
		\STATE Update constant background: $\hat{\b} \leftarrow \max\left(0, \frac{1}{T}(\mathbf{UV}-\hat{\mathbf{A}}\hat{\mathbf{C}})\mathbf{1}\right)$;
        \STATE$\mathbf{P} \leftarrow \left[\mathbf{V}^T,\mathbf{1}\right]\left(\left[\mathbf{U},-\b\right]^T\hat{\mathbf{A}}\right)$;
        \STATE$\mathbf{Q} \leftarrow \hat{\mathbf{A}}^{T}\hat{\mathbf{A}}$;
		\FOR{$k=1:K$}
		\STATE Update temporal: $\hat{\c}_{k} \leftarrow \max\left(0, \hat{\c}_{k} + \frac{\mathbf{P}(:,k)-\hat{\mathbf{C}}\mathbf{Q}(:,k)}{\mathbf{Q}(k,k)}\right)$;
		\ENDFOR
		\FOR{every 4 iterations}
		\FOR{$k=1:K$}
		\STATE $\mathrm{corr}(k,\x)\leftarrow \frac{1}{T\cdot\nu(\x)\cdot\mathrm{sd}(\c_k)}\mathbf{U}(\x,:)\left((\mathbf{V} - \bar{\mathbf{V}})(\c_k - \bar{\c}_k)\right)$;
		\STATE Update spatial support: $\Omega_k \leftarrow$ biggest connected component in $\{\x|\mathrm{corr}(k,\x)\geqslant\epsilon_1\}$ \\ that spatially overlaps with $\{\a_k>0\}$;
		\STATE $\hat{\a}_k(\Omega_k^{c}) \leftarrow 0$; 
		\STATE $\rho(k,\x)\leftarrow\left(\mathrm{corr}(k,\x)\geqslant\epsilon_2\right)$;
		\ENDFOR
		\STATE Merge overlapping components $k_1,k_2$ if $\sum_{\x} \left(\rho(k_1,\x) * \rho(k_2,\x)\right) / \sum_{\x}\rho(k_i,\x) \geqslant \epsilon_3$; 
		\STATE $(\tilde{\a},\tilde{\c}) \leftarrow$ rank-1 NMF on $[\hat{\a}_{k_1},\cdots,\hat{\a}_{k_r}][\hat{\c}_{k_1},\cdots,\hat{\c}_{k_r}]$ for merged components $k_1,\cdots,k_r$;
		\STATE $\hat{\mathbf{A}}\leftarrow \left[\hat{\mathbf{A}}\backslash \{\a_{k_1},\cdots,\a_{k_r}\},\tilde{\a}\right], \hat{\mathbf{C}}\leftarrow \left[\hat{\mathbf{C}}^T\backslash \{\c_{k_1},\cdots,\c_{k_r}\},\tilde{\c}\right]^T;$
		\STATE update number of components $K$; 
		\ENDFOR
		\ENDWHILE
		\RETURN $\hat{\mathbf{A}},\hat{\mathbf{C}},\hat{\b}$
	\end{algorithmic}
\end{algorithm}

\clearpage
\section*{Results}

\subsection*{Denoising}



\begin{table}[h!]
\centering
\begin{tabular}{cccccccc}
\toprule
\textbf{Dataset} & \multicolumn{3}{c}{\textbf{Dimensions}} & \textbf{Method}  & \textbf{Compression} & \textbf{Total} & \textbf{SNR}\\
\cmidrule{2-4}
 & Frames & FOV & Patch &  & \textbf{ratio} & \textbf{runtime (s)} & \textbf{metric} \\
\midrule
Endoscopic & 6000 & 256x256 & 16x16 & Patch-wise PMD  & 23  & 220.4 & 2.3\\
           &      &         & 16x16 & Patch-wise PCA* & X   & X     & X\\
           &      &         & NA    & Standard PCA    & 2 & 595.5 & 1.3\\
\midrule
Dendritic & 1000 & 192x192 & 16x16 & Patch-wise PMD & 52   & 3.2  & 3.7\\
          &      &         & 16x16 & Patch-wise PCA & 32   & 1.2  & 2.5\\
          &      &         & NA    & Standard PCA   & 2 & 18.3 & 1.1\\
\midrule
Three-photon & 3650 & 160x240 & 20x20 & Patch-wise PMD & 94  & 12.4   & 1.8\\
             &      &         & 20x20 & Patch-wise PCA & 44  & 3.5    & 1.4\\
             &      &         & NA    & Standard PCA   & 2 & 187.2  & 1.0\\
\midrule
Widefield & 1872 & 512x512 & 32x32 & Patch-wise PMD & 298  & 12.5 & 3.5\\
          &      &         & 32x32 & Patch-wise PCA & 265  & 10.1 & 3.4\\
          &      &         & NA    & Standard PCA   & 10  & 80.1 & 1.6\\
\midrule
Voltage & 6834 & 80x800 & 40x40 & Patch-wise PMD & 180  & 30.5   & 2.8\\
        &      &        & 40x40 & Patch-wise PCA & 213  & 8.7    & 2.7\\
        &      &        & NA    & Standard PCA   & 8  & 185.1  & 1.0\\
\bottomrule
\end{tabular}
\caption{Summary of performance for PCA vs.\ PMD(TV,TF).  SNR metric: average ratio of denoised vs raw SNR, with average restricted to top 10\% of pixels with highest raw SNR (to avoid division by small numbers when calculating SNR ratios); an SNR metric of 1 indicates no improvement compared to raw data.  Compression ratio defined in the main text.  * denotes that the patch-wise PCA method left a significant amount of visible signal in the residual for this dataset, and therefore we did not pursue further comparisons of timing or the other statistics shown here.  To obtain optimistic results for the standard PCA baseline, runtimes are reported for a truncated SVD with prior knowledge of the number of components to select for each dataset (i.e., runtimes did not include any model selection steps for standard PCA). Results for patch-wise methods are reported for a single (non-overlapping) tiling of the FOV; note that total runtimes are reported (not runtimes per patch). All experiments were run using an Intel Core i7-6850K 6-core processor.}
\label{tab:pro_pro}
\end{table}


\begin{figure}[t!]
\centering
	\includegraphics[page=1,width=18cm,height=14cm]{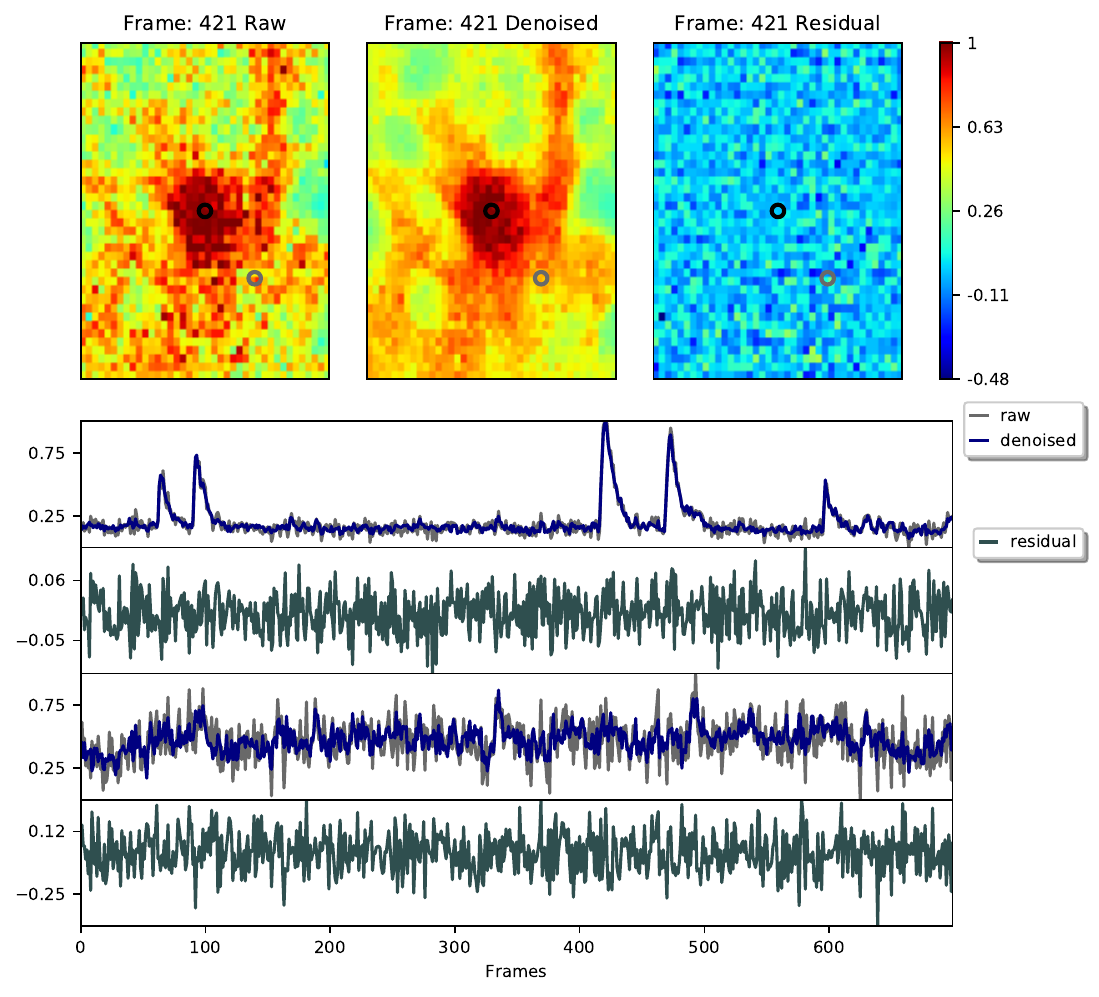}
    \caption{Illustration of the compression approach applied to microendoscopic imaging data.  Top: individual frame extracted from the raw movie $\mathbf{Y}$ (left), denoised movie $\hat{\mathbf{Y}}$ (middle), and residual $\mathbf{Y} - \hat{\mathbf{Y}}$ (right).  Bottom: example single-pixel traces from the movie (locations of pixels are circled in the top plots; first trace indicated by the black circle and second trace indicated by the gray circle).  Note that the denoiser increases SNR significantly, and minimal signal is left behind in the residual.  These results are best viewed in video form; see \href{\VideoEndoscopeURL}{microendoscopic imaging video} for details.}
    \label{fig:denoised_endoscope_1}
\end{figure}
\begin{figure}[t!]
\centering
	\includegraphics[page=2,width=18cm,height=14cm]{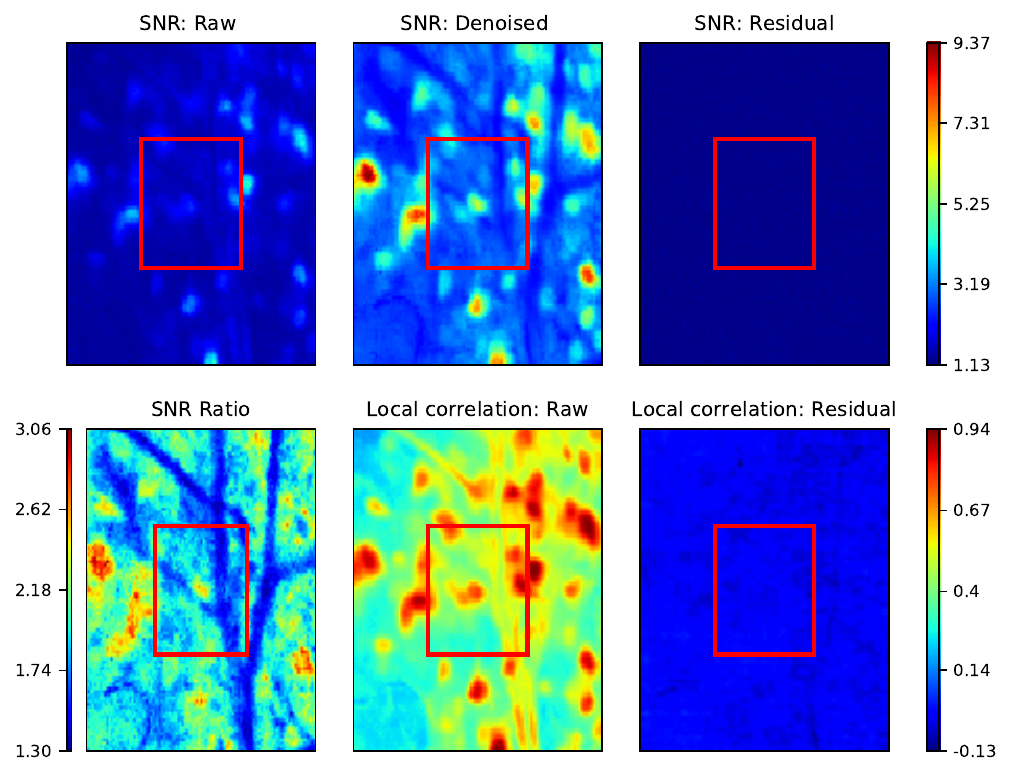}
    \caption{Further analysis of microendoscopic imaging data.  Top: per-pixel SNR estimated from the raw movie $\mathbf{Y}$ (left), denoised movie $\hat{\mathbf{Y}}$ (middle), and residual $\mathbf{Y} - \hat{\mathbf{Y}}$ (right).  Red box indicates zoomed-in region shown in the previous figure.  Bottom left panel: ratio of denoised vs.\ raw SNR; compression boosts SNR by roughly a factor of two here.  Bottom middle and right: ``correlation images" quantifying the average correlation of the temporal signals in each pixel vs.\ those in the nearest neighbor pixels \cite{Smith_2010}, computed on raw and residual data, indicating that minimal signal is left behind in the residual.  All results here and in the previous figure are based on background-subtracted data, for better visibility.
}
    \label{fig:denoised_endoscope_2}
\end{figure}
\begin{figure}[t!]
\centering
	\includegraphics[page=1,width=18cm,height=14cm]{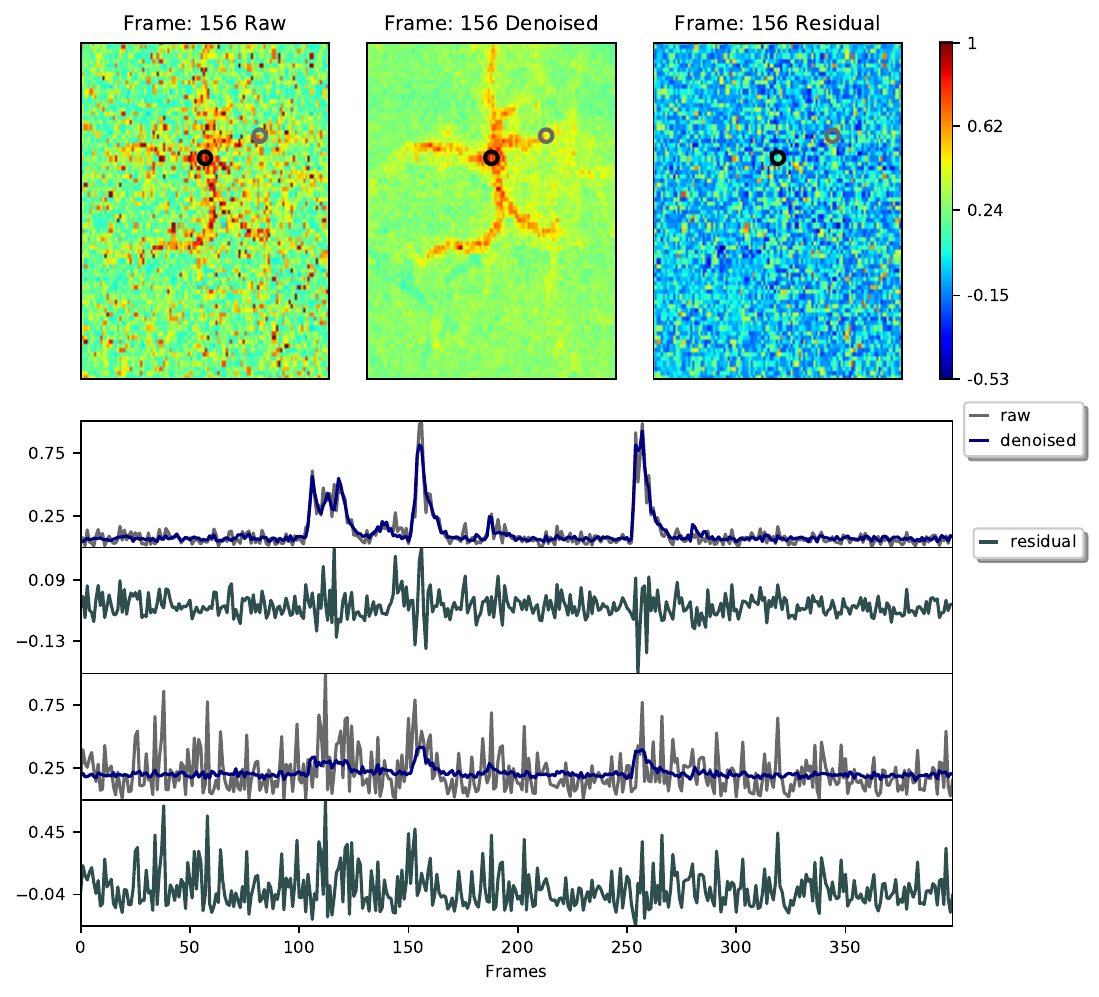}
    \caption{Example frames and traces from Bessel dendritic imaging data.  Conventions as in Figure \ref{fig:denoised_endoscope_1}. See \href{\VideoDemixDendriticURL}{Bessel dendritic imaging demixing video} for details.}
    \label{fig:denoised_dendritic_1}
\end{figure}
\begin{figure}[t!]
\centering
	\includegraphics[page=2,width=18cm,height=14cm]{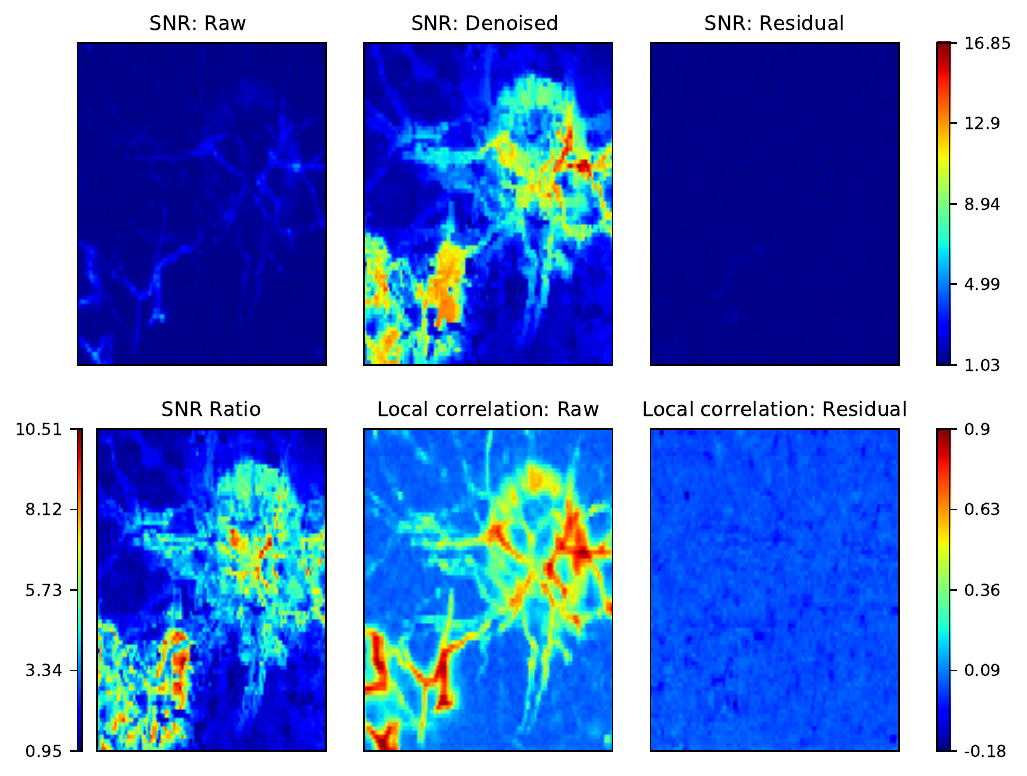}
    \caption{Summary quantification for denoising of Bessel dendritic imaging data.  Conventions as in Figure \ref{fig:denoised_endoscope_2}.
}
    \label{fig:denoised_dendritic_2}
\end{figure}


\begin{figure}[t!]
\centering
	\includegraphics[page=1,width=18cm,height=14cm]{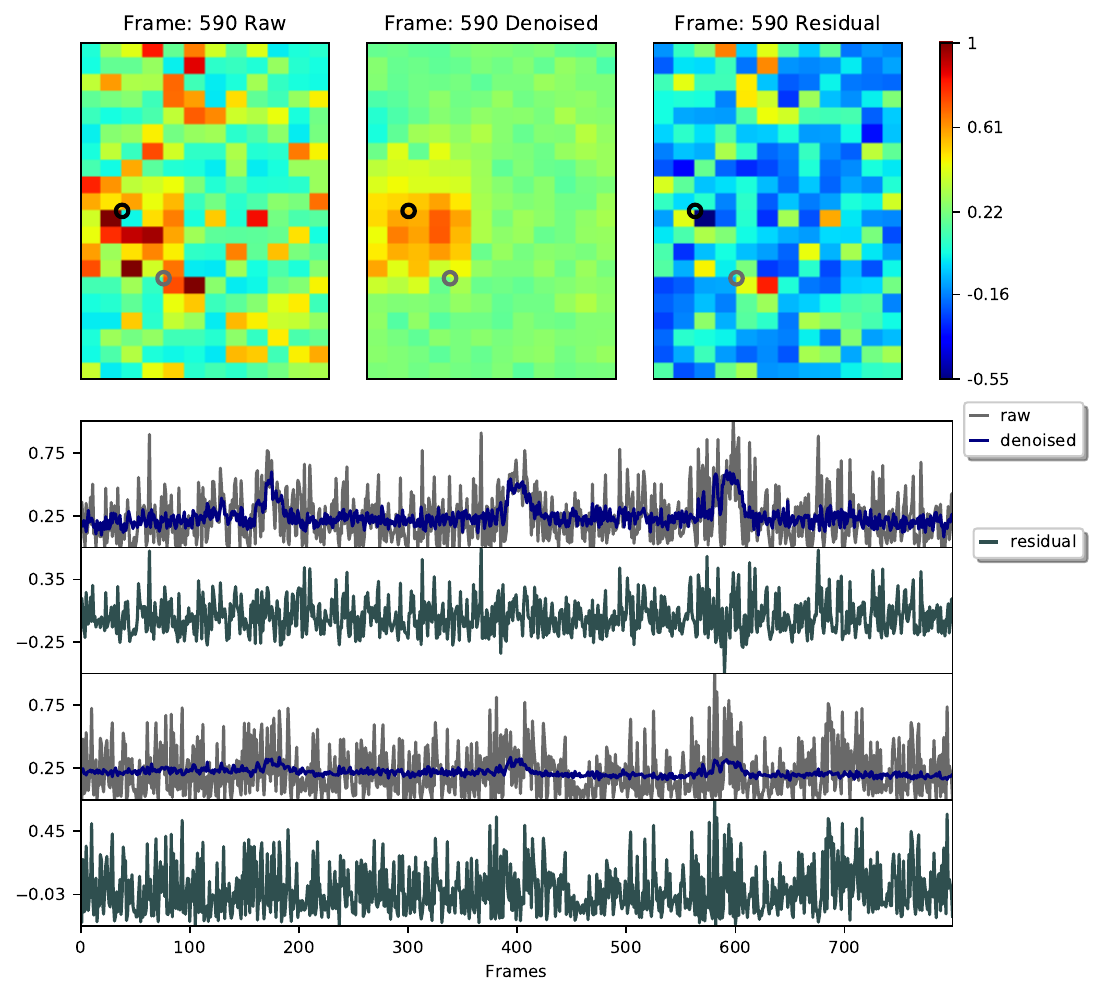}
    \caption{Example frames and traces from three-photon imaging data.  Conventions as in Figure \ref{fig:denoised_endoscope_1}.  See \href{\VideoThreePURL}{three-photon imaging video} for details.
}
    \label{fig:denoised_3p_1}
\end{figure}
\begin{figure}[t!]
\centering
	\includegraphics[page=2,width=18cm,height=14cm]{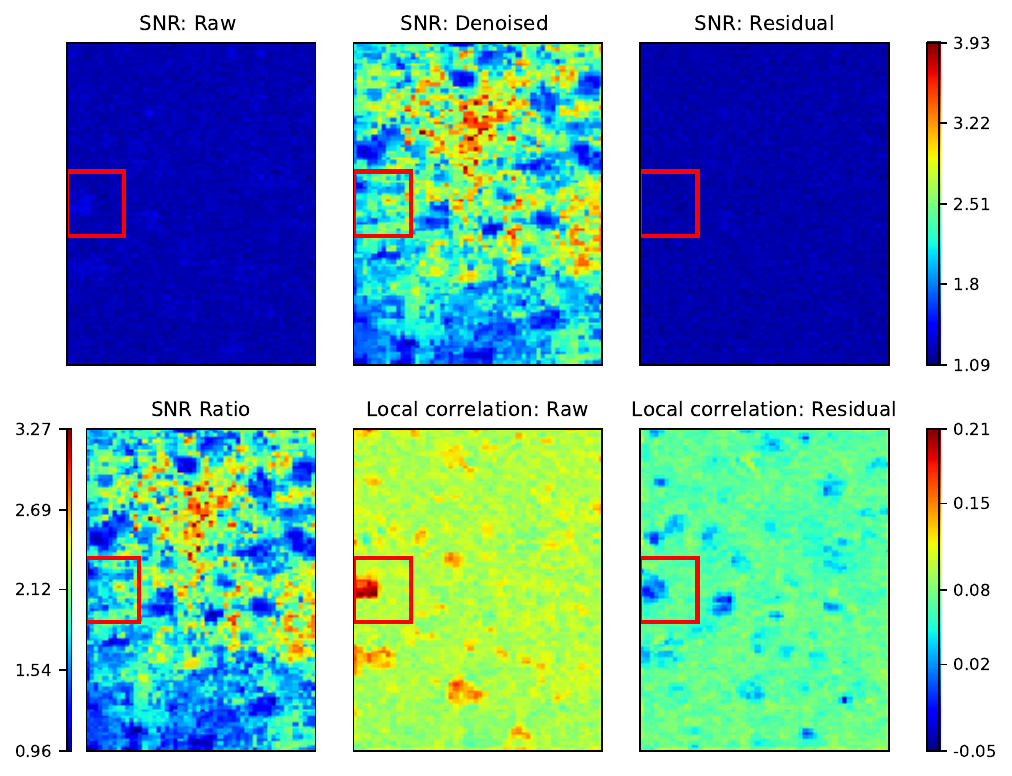}
    \caption{Summary quantification for denoising of three-photon imaging data.  Conventions as in Figure \ref{fig:denoised_endoscope_2}.
}
    \label{fig:denoised_3p_2}
\end{figure}
\begin{figure}[t!]
\centering
	\includegraphics[page=1,width=18cm,height=14cm]{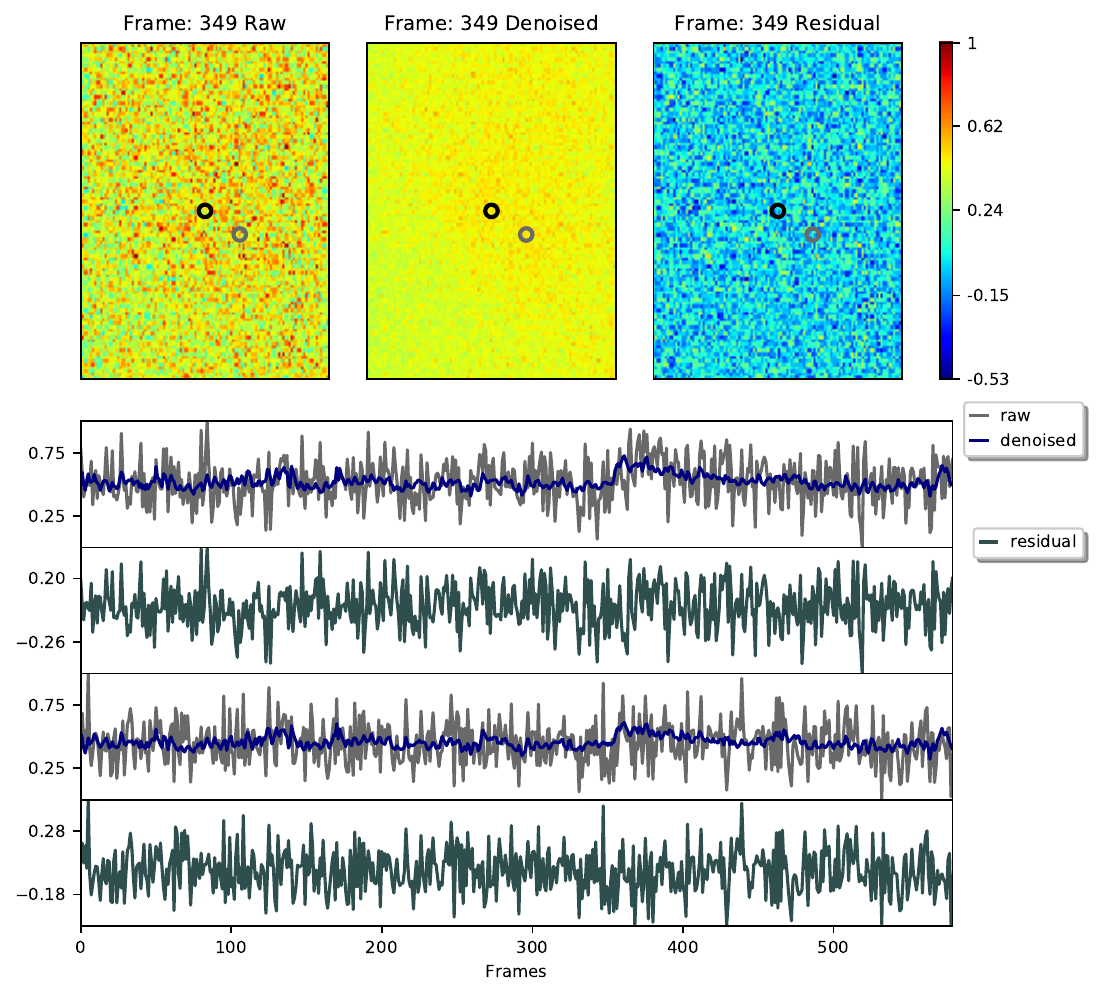}
    \caption{Example frames and traces from widefield imaging data.  Conventions as in Figure \ref{fig:denoised_endoscope_1}. See \href{\VideoWidefieldURL}{widefield imaging video} for details.
}
    \label{fig:denoised_widefield_1}
\end{figure}
\begin{figure}[t!]
\centering
	\includegraphics[page=2,width=18cm,height=14cm]{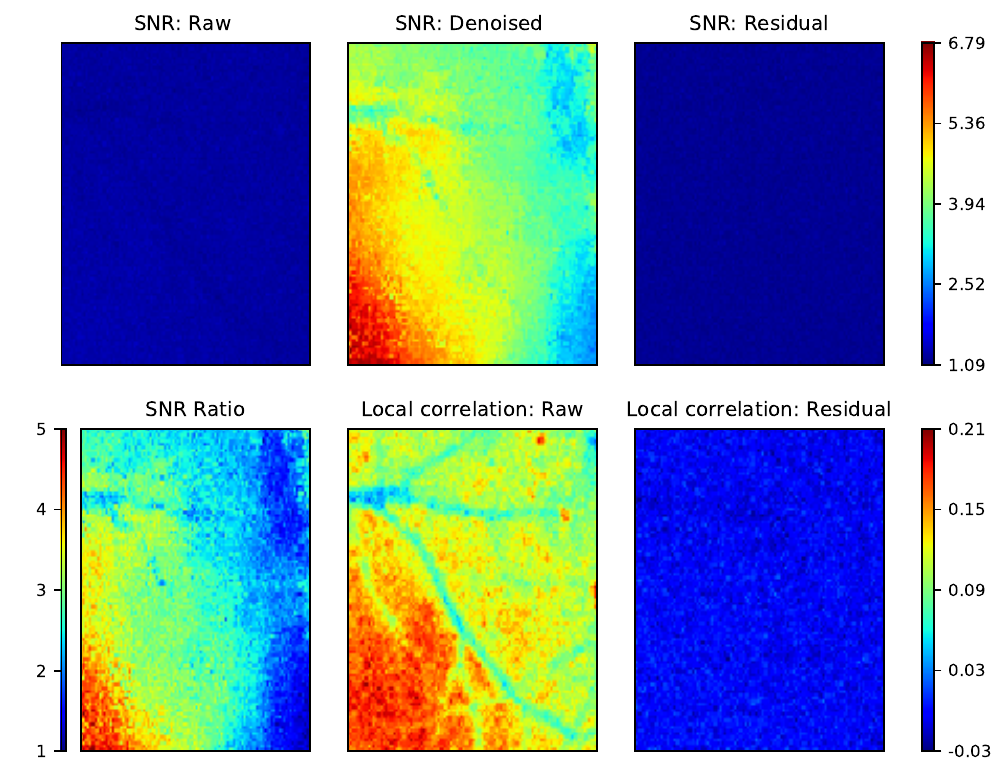}
    \caption{Summary quantification for denoising of widefield imaging data.  Conventions as in Figure \ref{fig:denoised_endoscope_2}.}
    \label{fig:denoised_widefield_2}
\end{figure}
\begin{figure}[t!]
\centering
	\includegraphics[page=1,width=18cm,height=14cm]{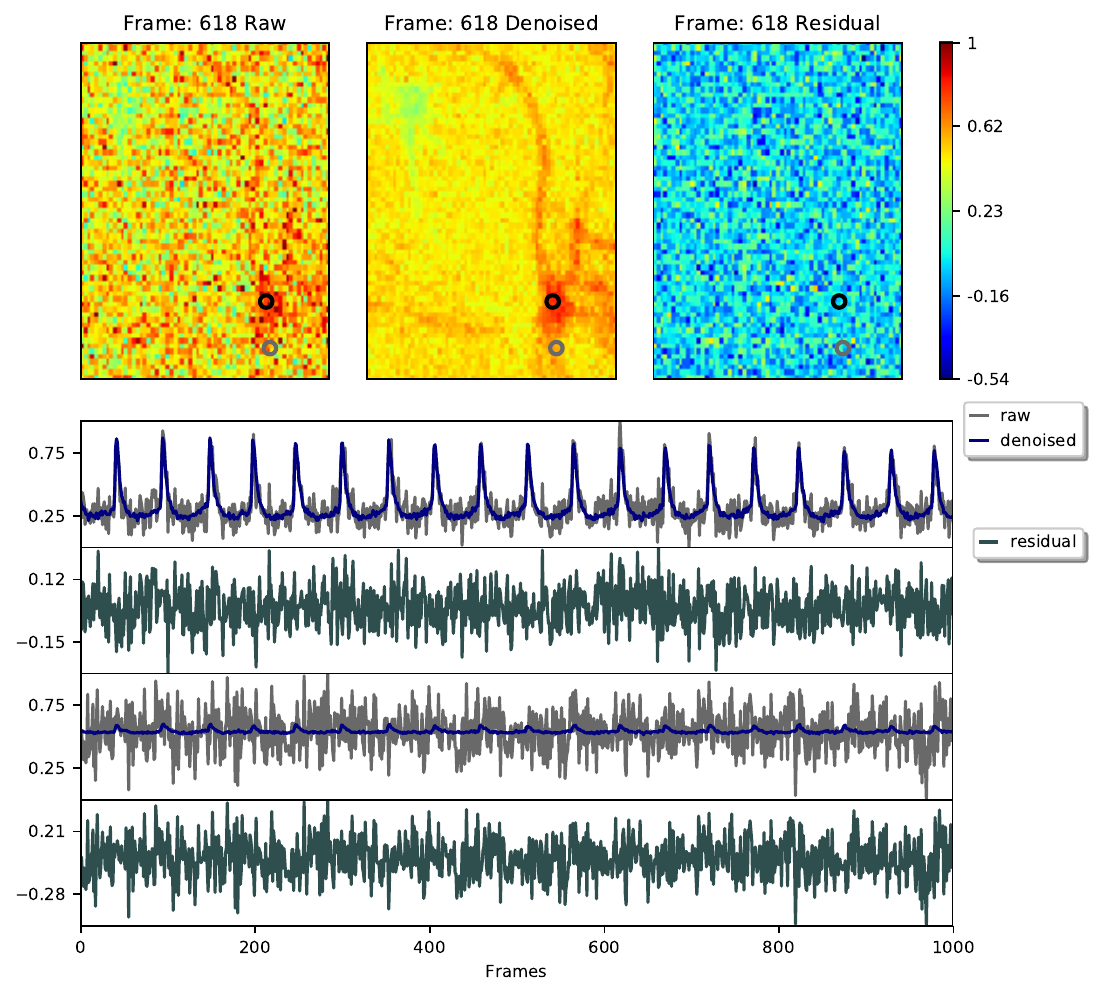}
    \caption{Example frames and traces from voltage imaging data. Conventions as in Figure \ref{fig:denoised_endoscope_1}. See \href{\VideoDemixVoltageURL}{voltage imaging demixing video} for details.}
    \label{fig:denoised_voltage_1}
\end{figure}
\begin{figure}[t!]
\centering
	\includegraphics[page=2,width=18cm,height=14cm]{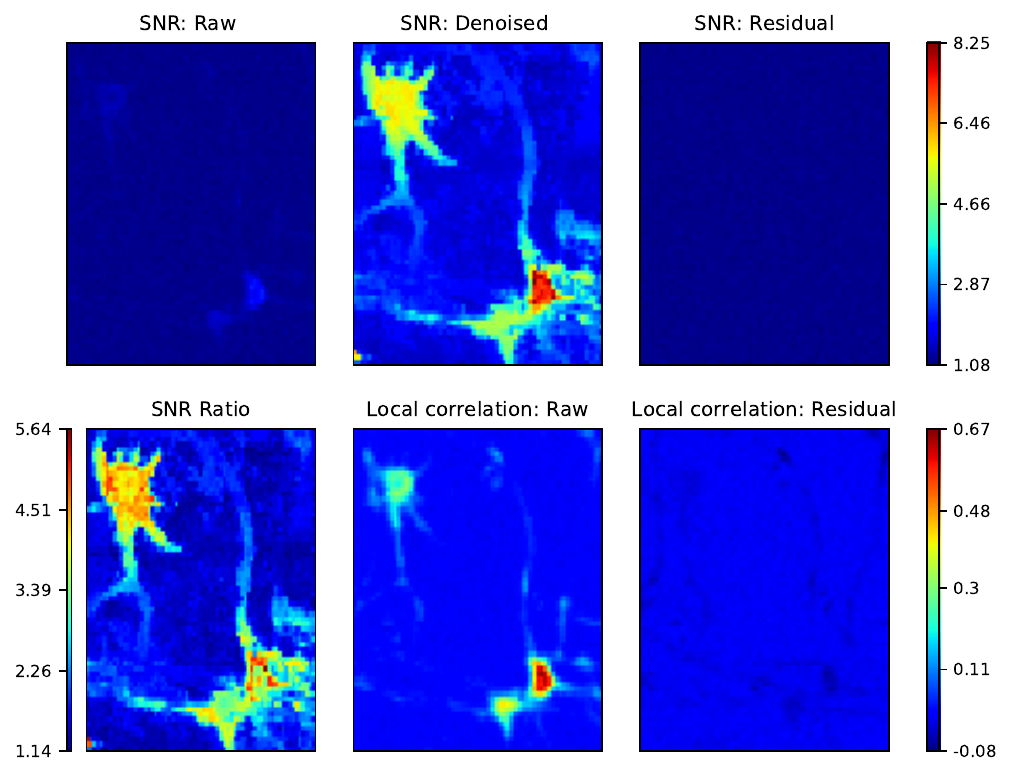}
    \caption{Summary quantification for denoising of voltage imaging data.  Conventions as in Figure \ref{fig:denoised_endoscope_2}.
}
    \label{fig:denoised_voltage_2}
\end{figure}

We have applied the denoising and compression approach described above to a wide variety of functional imaging datasets (See Appendix for full details):
\begin{itemize}
\item \textbf{Endoscopic}: one-photon microendoscopic calcium imaging in dorsal striatum of behaving mouse
\item \textbf{Dendritic}: two-photon Bessel-beam calcium imaging of dendrites in somatosensory cortex of mouse in vivo
\item \textbf{Three-photon}: three-photon calcium imaging of visual cortex of mouse in vivo
\item \textbf{Widefield}: one-photon widefield whole-cortex calcium imaging in behaving mouse
\item \textbf{Voltage}: one-photon in vitro voltage imaging under optogenetic stimulation.
\end{itemize}

The proposed methods perform well in all cases with no parameter tuning.  We obtain compression ratios (defined as $nnz(\mathbf{Y}) / [nnz(\mathbf{U})+nnz(\mathbf{V})]$, where $nnz(\mathbf{A})$ counts the number of nonzero elements of the matrix $\mathbf{A}$) of 20x-200x, and SNR improvements typically in the range of about 2x but ranging up to 10x, depending on the dataset and the region of interest (we find that SNR improvements are often largest in regions of strongest activity, so SNR improvements vary significantly from pixel to pixel).  See Table \ref{tab:pro_pro} and Figures \ref{fig:denoised_endoscope_1}-\ref{fig:denoised_voltage_2} for details.

In terms of runtime, we observed the expected scaling: the proposed method scales linearly in $T$, $d$, and the number of extracted components.  In turn, the number of estimated components scales roughly proportionally to the number of neurons visible in each movie (in the datasets with single-cell resolution).  Total runtimes ranged from a few seconds to a few minutes (for the ``Endoscope" dataset, which had the largest number of extracted components); these runtimes are fast enough for the proposed method to be useful as a pre-processing step to be run prior to  demixing.

We also performed comparisons against two simpler baselines: standard PCA run on the full dataset, and ``patch-wise PCA" run on the same patches as used by PMD.  For patch-wise PCA, we used the same stopping rule for choosing the rank of $\hat{\mathbf{Y}}$ as described above for PMD, but did not apply the TV or TF penalty.  We find that using the same rank selection criterion for PCA applied to the full dataset performs relatively poorly: in each of the five datasets examined, this approach left significant visible signal behind in the residual.  Thus, to make the comparisons as favorable as possible for standard PCA, we chose the rank manually, to retain as much visible signal as possible while keeping the rank as low as possible.  Nonetheless, we found that the PMD approach outperformed standard PCA significantly on all three metrics examined here (compression ratio, SNR improvement, and runtime), largely because PCA on the full image outputs dense $\mathbf{U}$ matrices (leading to slower computation and worse noise suppression) whereas the $\mathbf{U}$ matrices output by the patch-wise approaches are highly sparse.


The patch-wise PCA approach has much stronger performance than standard PCA applied to the full data.  In four out of five datasets (the ``Endoscope" dataset was the exception) patch-wise PCA captured all the visible signal in the dataset and did not leave any visible signal behind in the residual.  In these four datasets PMD performed comparably or significantly better than patch-wise PCA in terms of SNR improvement and compression score, but patch-wise PCA was faster.  Thus there may be some room to combine these two approaches, e.g., to use PCA as a fast initial method and then PMD to provide further denoising and compression.  We leave this direction for future work.

\clearpage

\subsection*{Demixing}

\subsubsection*{Voltage imaging data}

\begin{figure}[t!]
	\centering
	\begin{tabular}{rc}
		\ctab{A} & \includegraphics[width=1\textwidth,clip = true, trim = 5.cm 0.5cm -0.8cm 0.4cm]{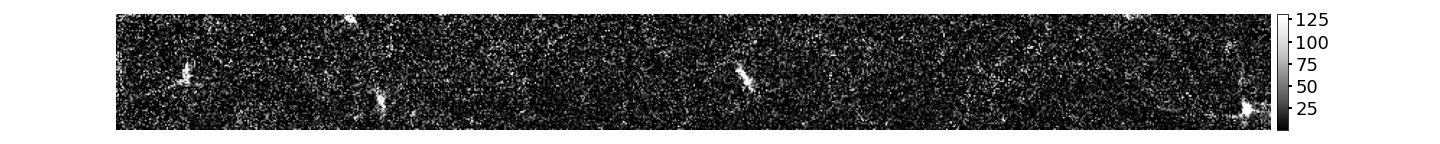}\\
		\ctab{B} & \includegraphics[width=1\textwidth,clip = true, trim = 5.cm 0.5cm -0.8cm 0.5cm]{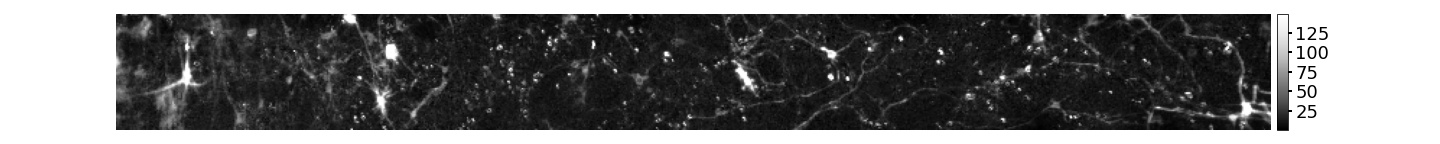}\\
		\ctab{C} & \includegraphics[width=1\textwidth,clip = true, trim = 5.5cm 0.5cm -1.5cm 0.5cm]{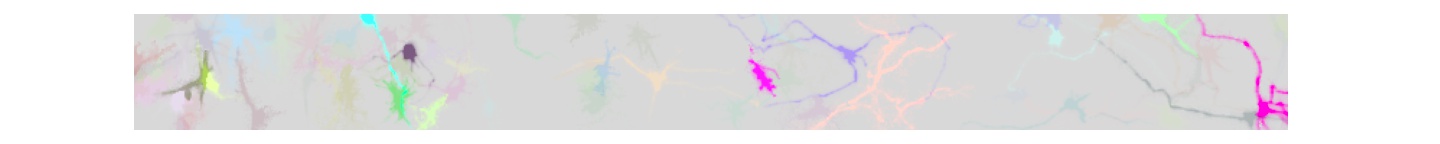}\\	
		\ctab{D} & \includegraphics[width=1\textwidth,clip = true, trim = 5.cm 0.5cm -1cm 0.4cm]{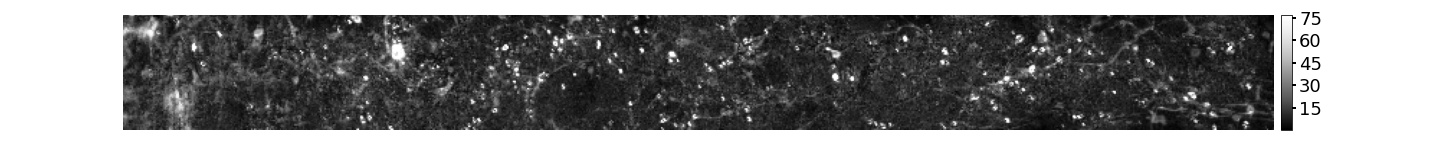}\\
		\ctab{E} & \includegraphics[width=1\textwidth,clip = true, trim = 4.8cm 0.5cm -0.7cm 0.5cm]{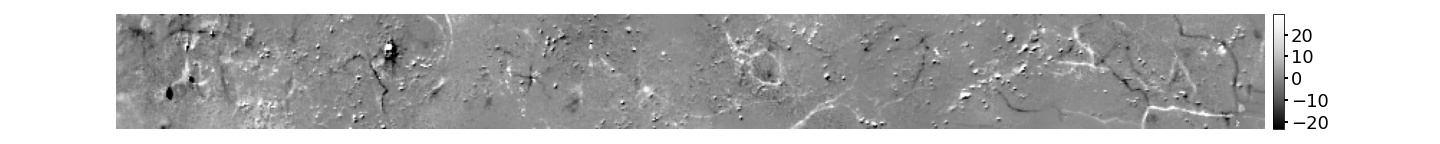}\\
		\ctab{F} & \includegraphics[width=1\textwidth,clip = true, trim = 5.4cm 0.5cm -0.5cm 0.5cm]{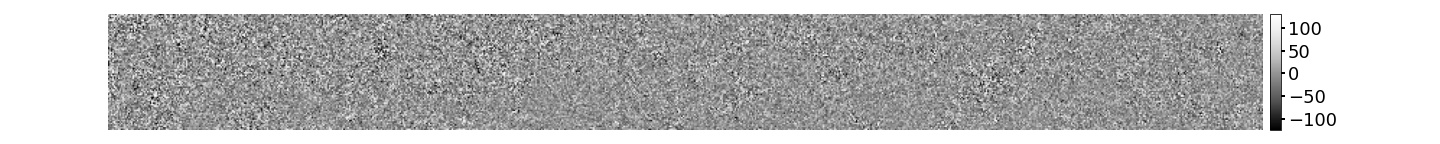}\\		
	\end{tabular}
\caption{An example frame illustrating demixing on voltage imaging data. (A) Detrended data $\mathbf{Y}$. (B) Denoised data $\hat{\mathbf{Y}}$. (C) Extracted signals $\mathbf{AC}$; each component $k$ is assigned a unique color, and the intensity of each pixel at each time is determined by the corresponding value of $\mathbf{AC}$.  (D) Constant background $\b$.  (E) Residual $\hat{\mathbf{Y}}-\mathbf{AC}-\b\mathbf{1}^T$.  (F) Noise removed in the denoising step.  See the \href{\VideoDemixVoltageURL}{voltage imaging demixing video} for a time-varying representation of the results here.} \label{fig:vi_demixing}
\end{figure}

\begin{figure}[t!]
	\centering
	\begin{tabular}{rllllll}
		\ctab{A} & \multicolumn{6}{l}{\includegraphics[width=1\textwidth,clip = true, trim = 5.6cm 0.5cm 0.5cm 0.5cm]{./plots/voltage/Yd_mean.jpg}}\\
		\ctab{B} & \multicolumn{6}{l}{\includegraphics[width=1\textwidth,clip = true, trim = 4.8cm 0.5cm -1.5cm 0.5cm]{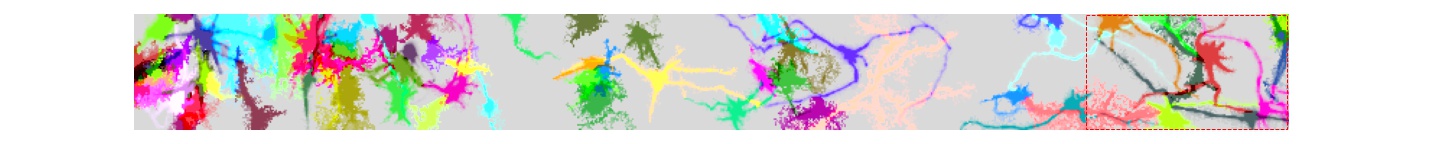}}\\
		\ctab{C} &\includegraphics[width=0.123\textwidth,clip = true, trim = 1.5cm 0.5cm 0cm 0.5cm]{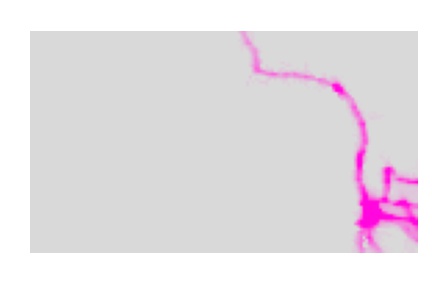}& \includegraphics[width=0.123\textwidth,clip = true, trim = 1.5cm 0.5cm 0.cm 0.5cm]{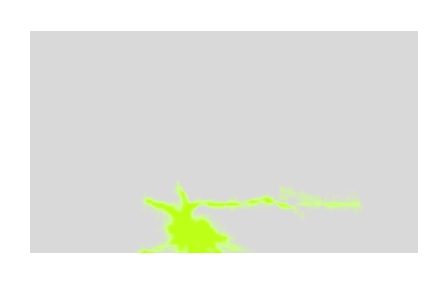} & {\includegraphics[width=0.123\textwidth,clip = true, trim = 1.5cm 0.5cm 0.cm 0.5cm]{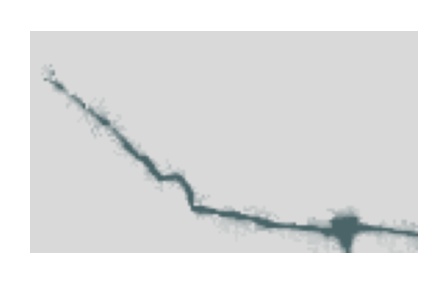}} & {\includegraphics[width=0.123\textwidth,clip = true, trim = 1.5cm 0.5cm 0.cm 0.5cm]{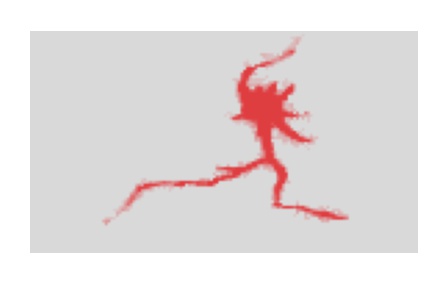}} & {\includegraphics[width=0.123\textwidth,clip = true, trim = 1.5cm 0.5cm 0.cm 0.5cm]{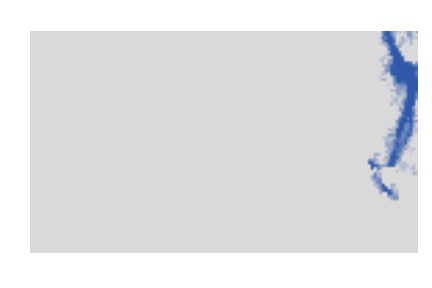}}&
{\includegraphics[width=0.123\textwidth,clip = true, trim = 1.5cm 0.5cm 0cm 0.5cm]{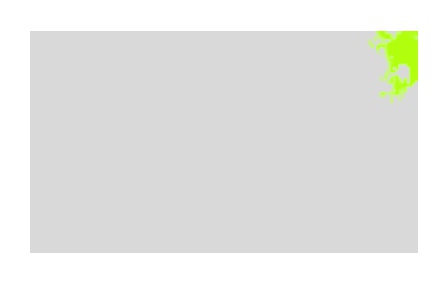}}\\
		\multirow{6}{*}{\ctab{D}} & \multicolumn{6}{l}{\includegraphics[width=1\textwidth,clip = true, trim = 0cm 0.7cm -5cm 0.6cm]{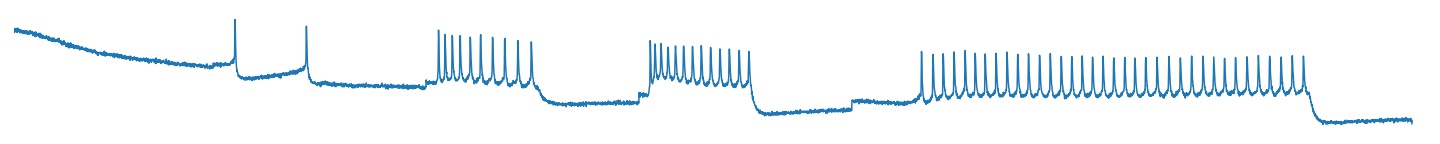}}\\	
		& \multicolumn{6}{l}{\includegraphics[width=1\textwidth,clip = true, trim = 0cm 0.7cm -5cm 0.6cm]{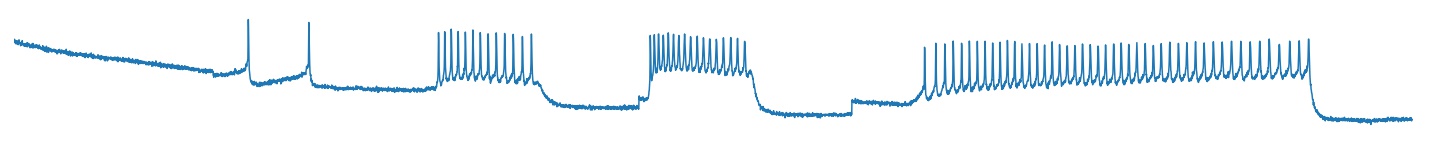}}\\	
		& \multicolumn{6}{l}{\includegraphics[width=1\textwidth,clip = true, trim = 0cm 0.7cm -5cm 0.6cm]{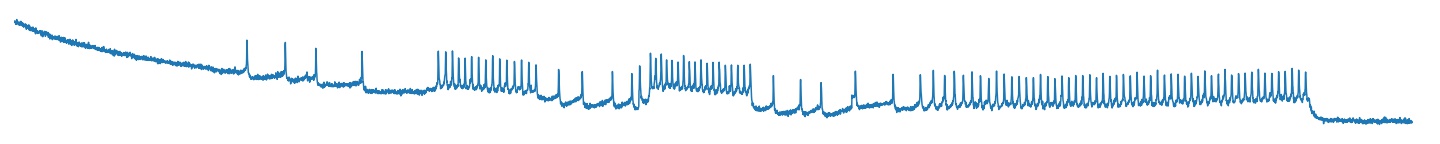}}\\
		& \multicolumn{6}{l}{\includegraphics[width=1\textwidth,clip = true, trim = 0cm 0.7cm -5cm 0.6cm]{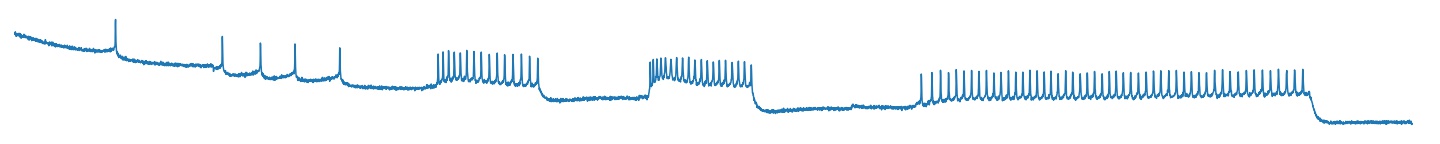}}\\	
		& \multicolumn{6}{l}{\includegraphics[width=1\textwidth,clip = true, trim = 0cm 0.7cm -5cm 0.6cm]{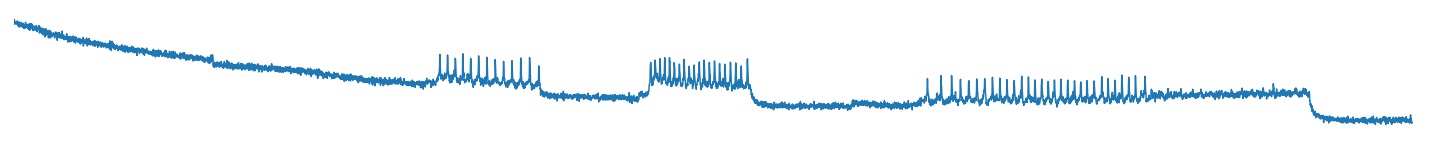}}\\	
		& \multicolumn{6}{l}{\includegraphics[width=1\textwidth,clip = true, trim = 0cm 0.5cm -5cm 0.6cm]{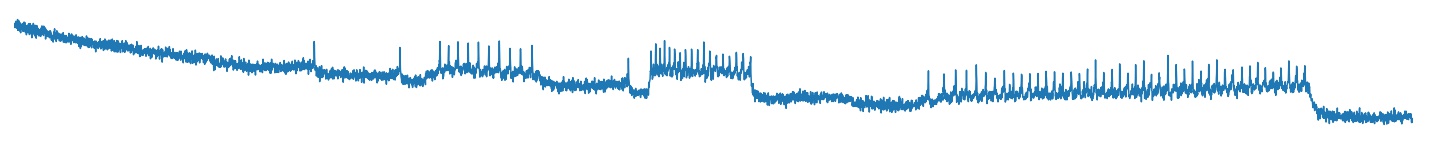}}\\
        \ctab{E} & \multicolumn{6}{l}{\includegraphics[width=1\textwidth,clip = true, trim = 0cm 0.5cm -5cm 0.6cm]{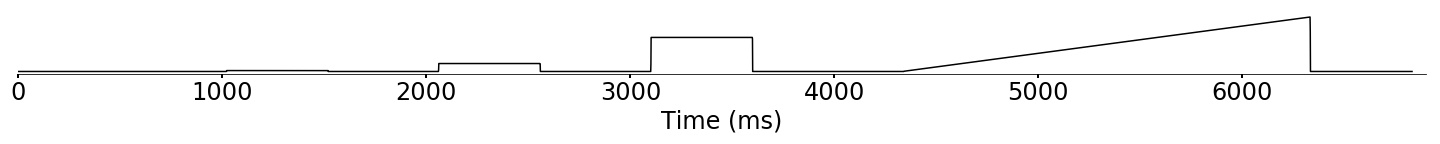}}\\
	\end{tabular}
\caption{Components extracted from voltage imaging data. (A) Mean intensity projection of $\hat{\mathbf{Y}}$.  (B) Extracted spatial components (each assigned a unique color).  (C) Details of the spatial components extracted in the zoomed-in patch (red outline in panel B), sorted in decreasing order of brightness.  (D) Raw temporal components corresponding to the spatial components shown in C (blue lines).  Note that the highly-correlated subthreshold activity and the strong bleaching trends visible in these components. (E) Optogenetic stimulation (consisting of three steps of increasing amplitude followed by a ramp; black line).} 
\label{fig:vi_components}
\end{figure}

Next we turn to the problem of demixing.  We begin with an analysis of a challenging voltage imaging dataset.  Voltage imaging (VI) data presents a few important challenges compared to calcium imaging (CI) data: currently-available VI data typically has much lower SNR and displays much stronger bleaching effects than CI data.  The dataset we focus on here has another challenging feature: the preparation was driven with time-varying full-field optogenetic stimulation, resulting in highly correlated subthreshold activity in the visible cells, which are highly overlapping spatially.  In preliminary analyses of this data we applied variants of CNMF-E \cite{zhou2018efficient} but did not obtain good results (data not shown), due to the strong bleaching and optogenetic stimulation-induced correlations present in this data.

Thus we pre-processed this data by applying a spline-based detrending to each pixel (see Appendix for full details).  This served to attenuate the highly-correlated bleaching signals and subthreshold fluctuations in the raw data, leaving behind spiking signals (which were not perfectly correlated at the millisecond resolution of the video data here) along with uncorrelated noise as the dominant visible signals in the data.  Figure \ref{fig:vi_superpixels} shows that the denoiser (followed by soft-thresholding) serves to significantly improve the separability of neural signals from noise in this data: the superpixels obtained after denoising and soft-thresholding provide excellent seeds for the constrained NMF analysis.  Figures \ref{fig:vi_demixing} (and the corresponding video) and \ref{fig:vi_components} demonstrate that the full demixing pipeline achieves good performance, extracting components with high spatial and temporal SNR and leaving relatively little residual signal behind despite the limited SNR and the multiple overlapping signals visible in the original (detrended) data.  Note that in the final step we project the estimated spatial components back onto the original data, recovering the (highly correlated) temporal components including strong bleaching components (panel D of Figure \ref{fig:vi_components}).  Finally, we achieved a speedup in the NMF iterations here that was roughly proportional to the ratio of the rank of $\mathbf{Y}$ compared to the rank of $\mathbf{U}$.


\clearpage

\subsubsection*{Bessel dendritic imaging data}

\begin{figure}[t!]
\centering
\begin{tabular}{ccc}
	\hspace{0.8cm} Proposed pipeline & \hspace{2.2cm} NMF on $\hat{\mathbf{Y}}$ & \hspace{-0.1cm} NMF on $\mathbf{Y}$\\
    \multicolumn{3}{c}{\includegraphics[width=1\textwidth,clip = true, trim = 1cm 1cm 0.cm 0.5cm]{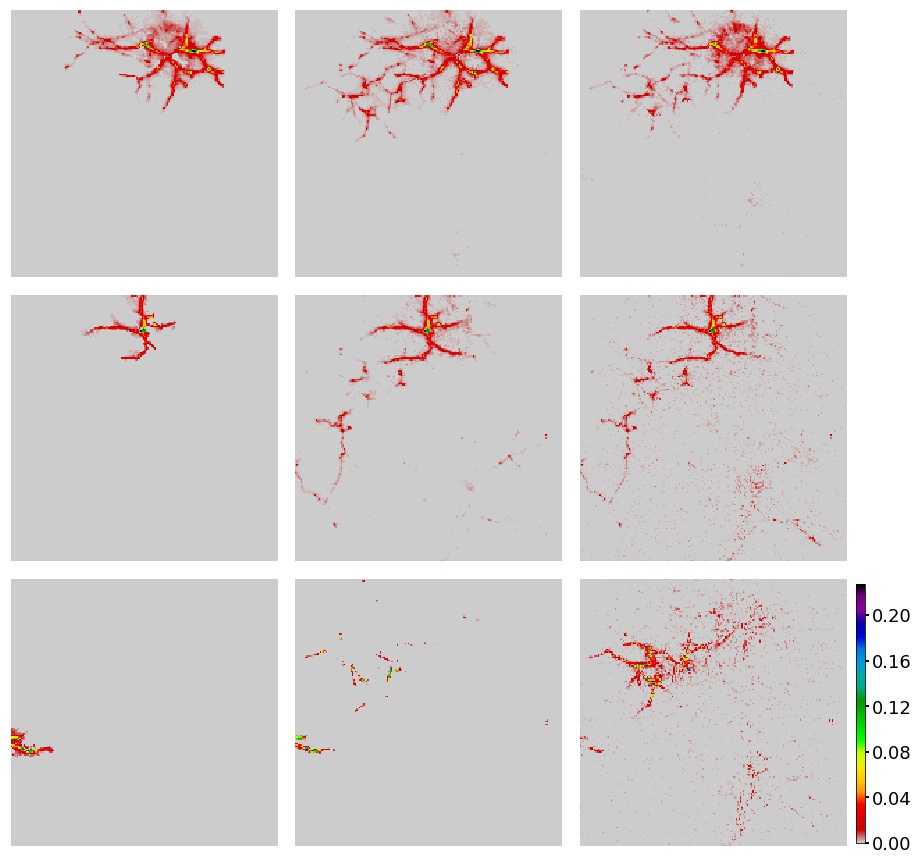}}
\end{tabular}
    \caption{Comparison of spatial components extracted from Bessel dendritic imaging data.  Each row shows best-matching components extracted by our proposed method (first column), sparse NMF on denoised data $\hat{\mathbf{Y}}$ (second column) and sparse NMF on raw data $\mathbf{Y}$ (third column). See the \href{\VideoDemixDendriticURL}{Bessel dendritic imaging demixing video} for further details.  The proposed pipeline extracts components that are significantly more localized and less noisy than the components extracted by sparse NMF; also note that denoising helps sparse NMF extract cleaner spatial components.}    \label{realcompare}
\end{figure}

\begin{figure}[t!]
\centering
\begin{tabular}{cccc}
	\hspace{0.55cm} Ground truth & \hspace{0.75cm} Proposed pipeline & \hspace{1.2cm} NMF on $\hat{\mathbf{Y}}$ & \hspace{-0.3cm} NMF on $\mathbf{Y}$\\
    \multicolumn{4}{c}{\includegraphics[width=1\textwidth,clip = true, trim = 1cm 1cm 0.cm 0.5cm]{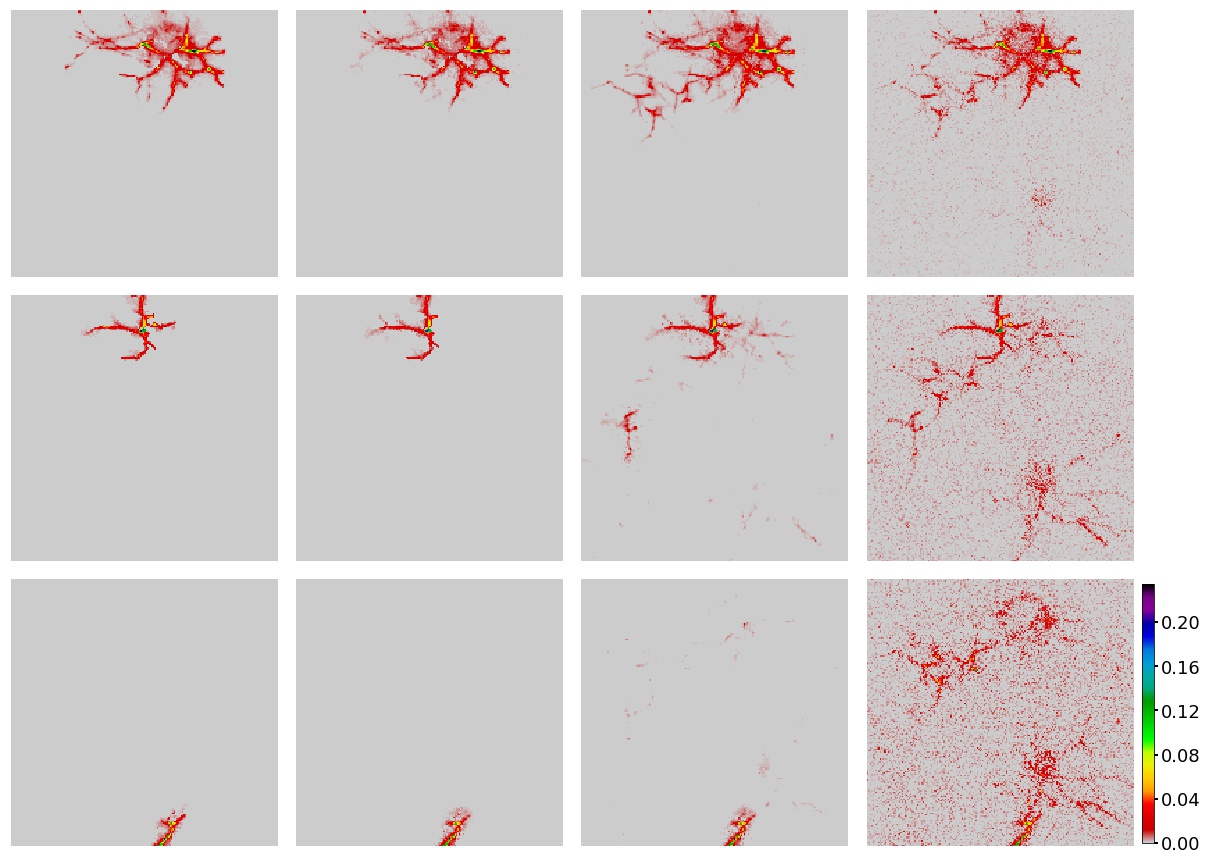}}
\end{tabular}
    \caption{Comparison to simulated ground truth based on Bessel dendritic imaging data.  Spatial components are arranged as in the previous figure, with the addition of ground truth components shown in the first column.  Note that the proposed pipeline recovers the ground truth simulated components much more accurately than do the sparse NMF baseline approaches.}    \label{simcompare}
\end{figure}

\begin{figure}[t!]
	\centering
	\begin{tabular}{ccc}
    \hspace{1.75cm} Spatial components & \hspace{0.85cm} Spatial components support & \hspace{-0.4cm} Temporal components\\
    \multicolumn{3}{c}{\includegraphics[width=1\textwidth,clip = true, trim = 0.2cm 0.3cm 0.2cm 0.2cm]{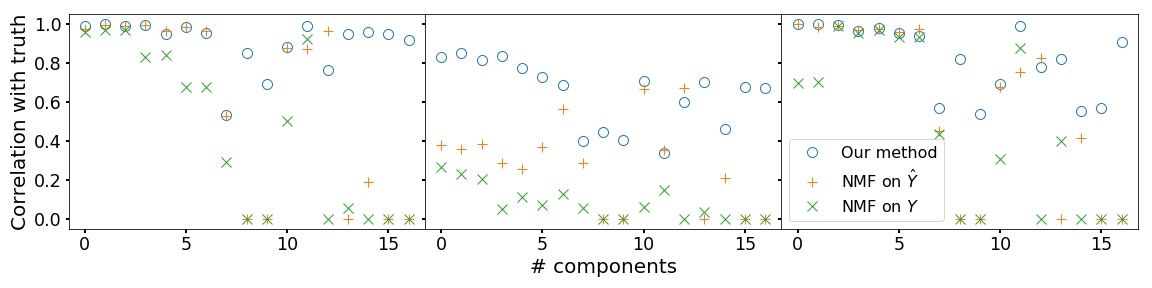}}\\
    \end{tabular}
    \caption{Quantification of comparisons on simulated Bessel dendritic imaging data.  Components are ordered by brightness; top 17 brightest components  shown here.  First column shows the correlation between true vs spatial components estimated by proposed pipeline (\textcolor{blue}{o}), sparse NMF on $\hat{\mathbf{Y}}$ (\textcolor{orange}{+}), and sparse NMF on $\mathbf{Y}$ (\textcolor{green}{x}).  Second column shows the correlation between the supports of the true and estimated spatial components.  Third column shows the correlation between the true vs estimated temporal components.  (The baseline NMF approaches missed some dimmer, weaker neurons, so the corresponding symbols are set to zero here.)  Note that components extracted by proposed pipeline typically have higher correlation with true components than sparse NMF baseline approaches.}
\label{corr_sim_comp}
\end{figure}

The VI dataset analyzed in the preceding subsection contained a number of large visible axonal and dendritic components, but also displayed strong somatic components.  For our next example we focus on a CI dataset dominated by dendritic components, where the simple Gaussian spatial filter approach introduced in \cite{pnevmatikakis2016simultaneous} for initializing somatic components is ineffective.  (Indeed, in dendritic or axonal imaging datasets, a search for ``hotspots" in the images is biased towards pixels summing activity from multiple neurons --- and these ``non-pure" pixels are exactly those we wish to avoid in the demixing initialization strategy proposed here.)

Figure \ref{realcompare} illustrates several of the spatial components extracted by our pipeline (again, see the corresponding video for a more detailed illustration of the demixing performance); these components visually appear to be dendritic segments and match well with the signals visible in the data movie.  Notably, no parameter tuning was necessary to obtain good demixing performance on both the VI and CI datasets, despite the many differences between these data types.  Additionally, as a baseline comparison we applied a simple sparse NMF approach with random initialization (similar to the method described in \cite{pnevmatikakis2016simultaneous}) to both the denoised and raw data ($\hat{\mathbf{Y}}$ and $\mathbf{Y}$, respectively). As shown in the right columns of Figure \ref{realcompare}, this baseline approach extracted components that were much more mixed and noisy than the components extracted by our proposed demixing pipeline; we also found that the baseline approach was more prone to missing weaker, dimmer components than was the proposed pipeline (data not shown).

The above analyses depended on qualitative visual examinations of the obtained components and demixing video.  We also generated simulated data with characteristics closely matched to the raw data, in order to more quantitatively test the demixing performance against a known (albeit simulated) ground truth.  To generate simulated data $\mathbf{Y}$, we used the $\mathbf{A}$ and $\mathbf{C}$ estimated from the raw data, and further estimated the conditional distribution of the residual as a function of the denoised data $\mathbf{A} \mathbf{C}$ in the corresponding pixel $x$ and time bin $t$;
then we added independent noise samples from this signal-dependent conditional distribution (but with the noise scale multiplied 2x, to make the simulation more challenging) to $\mathbf{AC}$. See the \href{\VideoSimulateDendriticURL}{simulated Bessel dendritic imaging video} for comparison of real and simulated data. We ran the three demixing pipelines on this simulated data.  Typical results of these simulations are shown in Figure \ref{simcompare}: again we see that the proposed pipeline captures the ground truth components much more accurately than do the baseline methods, similar to the results shown in Figure \ref{realcompare}.  Quantitatively, components extracted by proposed pipeline have higher correlation with ground truth components than do those extracted by sparse NMF approaches, as shown in Figure \ref{corr_sim_comp}. 


\section*{Discussion}


We have presented new scalable approaches for compressing, denoising, and demixing functional imaging data.  The compression and denoising methods presented are generally applicable and can serve as a useful generic step in any functional video processing pipeline, following motion correction and artifact removal.  The new demixing methods proposed here are particularly useful for data with many dendritic and axonal processes, where methods based on simple sparse NMF are less effective.







\subsection*{Related work}

Other work \cite{haeffele2014structured,pnevmatikakis2016simultaneous,de2018structured} has explored penalized matrix decomposition incorporating sparsity or total variation penalties in related contexts.  An important strength of our proposed approach is the focus on highly scalable patch-wise computations (similar to \href{https://github.com/flatironinstitute/CaImAn}{CaImAn}); this leads to fast computations and avoids overfitting by (implicitly) imposing strong sparsity constraints on the spatial matrix $\mathbf{U}$.  We also employ a constrained optimization approach using the trend-filtering (TF) penalty, which is more flexible e.g.~than the sparse convolutional temporal penalty used in \cite{haeffele2014structured}, since the constrained TF approach doesn't require us to fit a specific convolutional model or to estimate any Lagrange multipliers for the sparsity penalty.

There are also some interesting connections between the demixing approach proposed in \cite{petersen2017scalpel} and our approach to initializing NMF, which is based on the sparse projection algorithm (SPA).
\cite{fu2015self,gillis2018fast} discuss the relationships between SPA and group-sparse dictionary selection methods related to the approach used in \cite{petersen2017scalpel}; thus the methods we use to compute ``pure" superpixels and the methods used in \cite{petersen2017scalpel} to select neural dictionary elements are closely related.  However, our denoise-then-superpixelize approach to seeding the dictionary of neural temporal components is in a sense converse to the clustering approach developed in  \cite{petersen2017scalpel} for seeding the dictionary of neural spatial components.  There may be room to fruitfully combine these two approaches in the future.

\subsection*{Future work}

Real-time 
online updates for $\mathbf{U}$ and $\mathbf{V}$ should be possible, which would enable the incorporation of the compression and denoising approach into \cite{giovannucci2017onacid} for improved online demixing of neural activity.  We are also continuing to explore alternative methods for spatial and temporal denoising of $\mathbf{u}_k$ and $\mathbf{v}_k$, e.g.\ artificial neural network denoisers.

In the near future we plan to incorporate our code into the \href{https://github.com/flatironinstitute/CaImAn}{CaImAn} and \href{https://github.com/zhoupc/CNMF_E}{CNMF-E} packages for calcium imaging analysis.  We hope that the proposed compression methods will help facilitate more widespread and routine public sharing of these valuable datasets and lead to more open and reproducible neuroscience.

\subsection*{Code}

Open source code is available at \href{https://github.com/paninski-lab/funimag
}{https://github.com/paninski-lab/funimag}.

\subsection*{Video captions}

\begin{enumerate}
\item \hyperref{\VideoEndoscopeBkgURL}{}{pmd_videoendoscopebkg}{Microendoscopic imaging background video}\\
(left) Raw movie $\mathbf{Y}$; (middle) background $\mathbf{Y}_{BG}$ estimated via rank-5 PMD;  (right) estimated foreground $\mathbf{Y} - \mathbf{Y}_{BG}$.  Ticks along the horizontal and vertical axis (in this video and in the videos below) indicate patch borders; note that no edge artifacts are visible at these borders. 

\item \hyperref{\VideoEndoscopeURL}{}{pmd_videoendoscope}{Microendoscopic imaging video}\\
(left) Foreground; (middle) denoised foreground $\hat{\mathbf{Y}}$; (right) residual $\mathbf{Y} - \hat{\mathbf{Y}}$.

\item \hyperref{\VideoThreePURL}{}{pmd_video3p}{Three-photon imaging video}\\
(left) Raw movie $\mathbf{Y}$; (middle) denoised movie $\hat{\mathbf{Y}}$; (right) residual $\mathbf{Y} - \hat{\mathbf{Y}}$.

\item \hyperref{\VideoWidefieldURL}{}{pmd_videowidefield}{Widefield imaging video}\\
Same format as previous video.



\item \hyperref{\VideoSuperpixelizationURL}{}{sup_videosuperpixelization}{Superpixelization video}\\
Panels from top to bottom: (1) detrended movie $\mathbf{Y}$; (2) denoised movie $\hat{\mathbf{Y}}$; (3) MAD soft-thresholded movie; (4) rank-1 NMF approximation within superpixels; (5) superpixels; (6) pure superpixels.

\item \hyperref{\VideoDemixVoltageURL}{}{sup_videovoltage}{Voltage imaging demixing video}\\
Panels from top to bottom: (1) detrended movie $\mathbf{Y}$; (2) denoised movie $\hat{\mathbf{Y}}$; (3) estimated signal $\mathbf{AC}$; (4) background $\mathbf{B}$; (5) residual $\hat{\mathbf{Y}} - \mathbf{AC} - \mathbf{B}$; (6) estimated noise $\mathbf{Y} - \hat{\mathbf{Y}}$.

\item \hyperref{\VideoDemixDendriticURL}{}{sup_videodendritic}{Bessel dendritic imaging demixing video}\\
Top: (left) motion corrected movie $\mathbf{Y}$; (middle) denoised movie $\hat{\mathbf{Y}}$;  (right) estimated signal $\mathbf{AC}$;
Bottom: (left) background $\mathbf{B}$, (middle) residual $\hat{\mathbf{Y}} - \mathbf{AC} - \mathbf{B}$, and (right) estimated noise $\mathbf{Y} - \hat{\mathbf{Y}}$.

\item \hyperref{\VideoSimulateDendriticURL}{}{sup_videosimulatedendritic}{Simulated Bessel dendritic imaging video}\\
Top: (left) Motion corrected real movie; (right) simulated movie.
Bottom: (left) estimated noise from real movie; (right) simulated noise.
\end{enumerate}

\subsection*{Acknowledgments}

We thank Shay Neufeld and Bernardo Sabatini for generously sharing their micro-endoscopic data with us, and Andrea Giovanucci, Eftychios Pnevmatikakis, Ziqiang Wei, Darcy Peterka, Jack Bowler, and Uygar Sumbul for helpful conversations.  We also thank our colleagues in the International Brain Laboratory for motivating our efforts towards compressing functional imaging data.  This work was funded by Army Research Office W911NF-12-1-0594 (MURI; EH and LP), the Simons Foundation Collaboration on the Global Brain (LP), National Institutes of Health R01EB22913 (LP), R21EY027592 (LP), 1U01NS103489-01 (NJ and LP), R01NS063226 (EH), R01NS076628 (EH), RF1MH114276 (EH), and U19NS104649-01 (EH and LP); in addition, this work was supported by the Intelligence Advanced Research Projects Activity (IARPA) via Department of Interior/ Interior Business Center (DoI/IBC) contract number D16PC00003 (LP). The funders had no role in study design, data collection and analysis, decision to publish, or preparation of the manuscript.

\subsection*{Author contributions}

EKB and LP conceived the project.  EKB led development of the local PCA compression and denoising approach, including the 4x overcomplete approach for avoiding block artifacts.  IK led development of the PMD(TF,TV) approach.  DZ led development of the superpixelization and local NMF demixing approach.  RZ developed a preliminary version of the PMD approach.  PZ contributed to the development of the demixing approach.  FG, JF, and GD contributed the voltage imaging dataset.  JR, PF, TM, and AT contributed the three-photon imaging dataset.  YL, RL, and NJ contributed the Bessel dendritic dataset.  YM, SK, MS, and EH contributed the widefield dataset.  EKB, IK, DZ, and LP wrote the paper, with input from PZ.  LP supervised the project.

\appendix

\section*{Appendix: dataset details}
\subsection*{Microendoscopic imaging data}

This dataset was analyzed previously in
\cite{zhou2018efficient}; see the ``Dorsal Striatum Data" subsection of the Methods section of that paper for full experimental details.  Briefly, a 1 mm gradient index of refraction (GRIN) lens was implanted into dorsal striatum of a mouse expressing AAV1-Syn-GCaMP6f; imaging was performed using a miniature one-photon microscope with an integrated 475 nm LED (Inscopix) while the mouse was freely moving in an open-field arena.  Images were acquired at 30 Hz and then down sampled to 10 Hz.

\subsection*{Bessel dendritic imaging data}

All surgical procedures were in accordance with protocols approved by the Howard Hughes Medical Institute Janelia Research Campus Institutional Animal Care and Use Committee. C57BL/6J mice over 8 weeks old at the time of surgery were anesthetized with isoflurane anesthesia (1–2\%). A craniotomy over nearly the entire left dorsal cortex (from Bregma +3 mm to Bregma -4.0 mm) was performed with the dura left intact, with the procedure described in detail previously in \cite{sofroniew2016large}. AAV2/9-synapsin-flex-GCaMP6s (2.5$\times 10^{13}$ GC/ml) was mixed with AAV2/1-synapsin-Cre (1.5$\times 10^{13}$ GC/ml, 1000$\times$dilution with PBS) at 1:1 to make the working viral solution for intracerebral injections.  30 nl viral solution was slowly injected into exposed cortex at 0.5 mm below dura. Injection sites were evenly spaced (at 0.7-0.9 mm separation) along two lines at 2.3 mm and 3.3 mm parallel to the midline. A custom-made glass coverslip (450 $\mu$m thick) was embedded in the craniotomy and sealed in place with dental acrylic. A titanium head bar was attached to the skull surrounding the coverslip. After recovery from surgery, the mice were habituated to head fixation. Four weeks after surgery, the head-fixed mouse was placed on a floating ball in the dark. The spontaneous neural activity as indicated by GCaMP6s fluorescence signal was recorded in the somatosensory cortex. 

\begin{figure}[t!]
\centering
\includegraphics[width=1\textwidth]{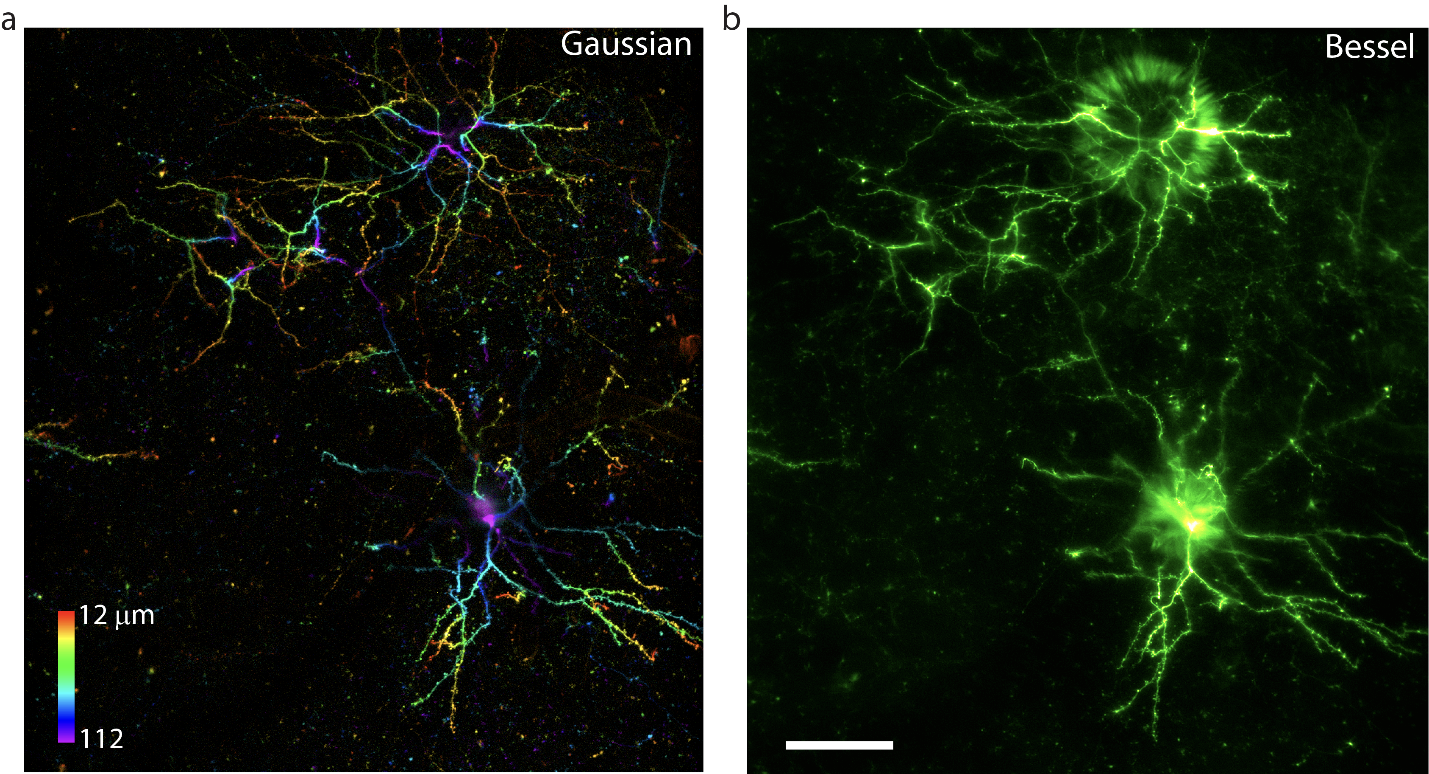}
\caption{In vivo volumetric imaging of dendrites in the mouse brain. (a) Maximum intensity projection of a 3D volume (635 $\mu$m x 694 $\mu$m x 100 $\mu$m) of dendrites. The sampling size was 0.33 $\mu$m/pixel. Post-objective power: 24 mW. (b) Image of the same volume collected by scanning a Bessel focus with 0.60 $\mu$m lateral FWHM and 71 $\mu$m axial FWHM. The effective volume rate was 3.7 Hz. Post-objective power: 120 mW. Excitation wavelength: 970 nm. Scale bar: 100 $\mu$m. 
}\label{gaussian_bessel}
\end{figure}

Volumetric imaging of dendrites was achieved by scanning an axially extended Bessel focus in \cite{lu201850} and \cite{lu2017video}. An axicon-based Bessel beam module was incorporated into a 2-photon random access mesoscope (2p-RAM) in \cite{lu201850}. Details of the 2p-RAM have been described previously in \cite{sofroniew2016large}. Briefly, the system was equipped with a 12kHz resonant scanner (24 kHz line rate) and a remote focusing unit that enabled fast axial movements of the focal plane. The system has an excitation numerical aperture (NA) of 0.6 and a collection NA of 1.0. The measured lateral full width at half maximum (FWHM) of the Gaussian focus at the center of the field of view was ~0.65 $\mu$m. The lateral and axial FWHMs of the Bessel focus were 0.60 $\mu$m and 71 $\mu$m, respectively. Scanning the Bessel focus in two dimensions, therefore, probed brain volumes within a ~100 $\mu$m axial range. The volumetric dendritic data presented in this paper were obtained by placing the center of the Bessel focus at 62 $\mu$m below dura to probe structures at 12 $\mu$m to 112 $\mu$m below dura (figure \ref{gaussian_bessel}). Dendrites within this volume were imaged at an effective volume rate of 3.7 Hz, with each image having 1924$\times$2104 pixels at 0.33 $\mu$m/pixel in the x-y plane. The wavelength of the excitation light was 970 nm and the post-objective excitation power was 120 mW. Images were spatially decimated and cropped for the analyses shown here.

\subsection*{Three-photon imaging data}

All procedures were carried out in accordance with the ethical guidelines of the National Institutes of Health and were approved by the Institutional Animal Care and Use Committee (IACUC) of Baylor College of Medicine. Cranial window surgeries over visual cortex were performed as described previously \cite{reimer2014pupil}. Briefly, a 4 mm cranial window was opened under isoflurane anesthesia and sealed with a 4 mm glass coverslip and surgical glue. The dura was removed before applying the coverslip to increase optical access to the cortex. Imaging was performed in a triple-transgenic mouse (Slc17a7-Cre x Dlx5-CreER x Ai148) expressing GCaMP6f pan-neuronally throughout cortex. Three-photon imaging data was collected as described previously \cite{ouzounov2017vivo}. Three-photon excitation of GCaMP6 was at 1320nm, which also enabled visualization of unlabeled vasculature and white matter via THG (third harmonic generation). Power was calibrated prior to each day of scanning and carefully maintained below 1.5nJ at the focal plane. For this study, scans were collected at 680 microns and 710 microns below the cortical surface with a 540 x 540 micron field of view at 0.59 pixels/micron spatial resolution and a frame rate of 5 Hz. Imaging was performed at the border of V1 and LM during presentation of oriented noise stimuli.

\subsection*{Widefield imaging data}

See \cite{ma2016resting,ma2016wide} for full details.

\subsection*{Voltage imaging data}

Q-State's proprietary Optopatch all-optical electrophysiology platform was used to record fluorescence recordings from induced pluripotent stem (iPS) cell-derived NGN2 excitatory neurons from a cohort of human subjects \cite{werley2017all}. Stimulation of action potentials was achieved with a blue light-activated channelrhodopsin (CheRiff). Fluorescent readout of voltage was enabled by an Archaerhodopsin variant (QuasAr). NGN2 neurons were produced at Q-State using a transcriptional programming approach. Recordings were performed with an ultra-widefield instrument with a resolution of 800x80 pixels (corresponding field of view of 2 mm$^2$) at a frame rate of 987 Hz.

The obtained data displayed breaks during stimulus resets and photobleaching.  To remove these effects from the raw data, we removed frames during stimulus resets, extracted slow trends with a robust B-spline regression (with knots chosen to allow for non-differentiability at stimulus change-points and discontinuity at stimulus resets), and then a quadratic regression against frames with no stimuli to capture and then remove photobleaching effects.

\clearpage

\bibliographystyle{apalike}
\bibliography{axon_pipeline}
\end{document}